\documentclass[pdflatex,sn-mathphys-num]{sn-jnl}
\usepackage{siunitx}

\usepackage{graphicx}%
\usepackage{multirow}%
\usepackage{amsmath,amssymb,amsfonts}%
\usepackage{amsthm}%
\usepackage{mathrsfs}%
\usepackage[title]{appendix}%
\usepackage{xcolor}%
\usepackage{textcomp}%
\usepackage{manyfoot}%
\usepackage{booktabs}%
\usepackage{algorithm}%
\usepackage{algorithmicx}%
\usepackage{algpseudocode}%
\usepackage{listings}%
\usepackage{lmodern}

\begin{document}

\title[Article Title]{Meta-heuristic Hypergraph-Assisted Robustness Optimization for Higher-order Complex Systems}

\author[1]{\fnm{Xilong} \sur{Qu}}\email{quxilong@mail.dlut.edu.cn}

\author*[1]{\fnm{Wenbin} \sur{Pei}}\email{peiwenbin@dlut.edu.cn}

\author[1]{\fnm{Haifang} \sur{Li}}\email{lihf@mail.dlut.edu.cn}

\author*[1]{\fnm{Qiang} \sur{Zhang}}\email{zhangq@dlut.edu.cn}

\author[2]{\fnm{Bing} \sur{Xue}}\email{bing.xue@ecs.vuw.ac.nz}

\author[2]{\fnm{Mengjie} \sur{Zhang}}\email{mengjie.zhang@ecs.vuw.ac.nz}

\affil*[1]{\orgdiv{School of Computer Science and Technology}, \orgname{Dalian University of Technology}, \orgaddress{\street{Lingshui Street}, \city{Dalian}, \postcode{116024}, \state{Liaoning}, \country{China}}}

\affil*[2]{\orgdiv{School of Engineering and Computer Science}, \orgname{Victoria University of Wellington}, \orgaddress{\street{PO Box 600}, \city{Wellington}, \postcode{6140}, \country{New Zealand}}}



\abstract{In complex systems (e.g., communication, transportation, and biological networks), high robustness ensures sustained functionality and stability even when resisting attacks. 
However, the inherent structure complexity and the unpredictability of attacks make robustness optimization challenging. Hypergraphs provide a framework for modeling complicated higher-order interactions in complex systems naturally, \textcolor{black}{but their potential has not been systematically investigated.} Therefore, we propose an effective method based on genetic algorithms from Artificial Intelligence to optimize the robustness of complex systems modeled by hypergraphs. 
By integrating percolation-based metrics with adaptive computational techniques, our method \textcolor{black}{achieves improved accuracy and efficiency}. Experiments on both synthetic and real-world hypergraphs demonstrate the effectiveness of the proposed method in mitigating malicious attacks, with robustness improvements ranging from 16.6\% to 205.2\%. Further in-depth analysis reveals that optimized hypergraph-based systems exhibit a preferential connection mechanism in which high-hyperdegree nodes preferentially connect to lower-cardinality hyperedges, forming a distinctive Lotus topology that significantly improves robustness. Based on this finding, we propose a robust hypergraph generation method that allows robustness to be controlled via a single parameter $rb$. Notably, for $rb<-1$, a distinct Cactus topology emerges as an alternative to the Lotus topology observed for $rb>1$. The discovery of the Lotus and Cactus topologies offers valuable insights for designing robust higher-order networks while providing a useful foundation for investigating cascading failure dynamics in complex systems.}




\maketitle




The system robustness \cite{roberts1980robustness}, normally defined as the capacity to maintain functionality and performance under internal failures, external disturbances, or environmental fluctuations, \textcolor{black}{is critical to} maintain the system stability \cite{meena2023emergent,kiss2024production,sun2024global}. Complex networks \cite{newman2003structure} are the most representative complex systems, which capture the intricate relationships between their constituent components. The network robustness manifests most critically in infrastructure systems where failure cascades pose existential risks, e.g., power grids \cite{andersson2005causes,dobson2007complex}, communication networks \cite{falkenberg2024patterns}, and transportation networks \cite{scagliarini2025assessing}. For example, a robust power grid \cite{yang2017small} is essential to ensure a reliable energy supply, whereas robust communication networks \cite{tu2000robust} underpin the seamless exchange of information on a global scale.


Recent studies \cite{reis2014avoiding,schneider2011mitigation,huang2025enhancing,wang2019surrogate,zhou2023combined} highlighted the critical impact of network structure on robustness \textcolor{black}{for complex systems}, driving significant research efforts toward its optimization.
However, enhancing robustness through structure optimization remains a significant challenge, due mainly to the following two difficulties. First, the large number of nodes and edges in large-scale networks \cite{dong2024survey} leads to an exceptionally complex optimization space. Second, the presence of intricate higher-order relationships among nodes \cite{battiston2020networks}, coupled with the highly unpredictable nature of node failures \cite{liu2023threshold}, further complicates the optimization process. 
Node failures can be broadly classified into two categories: random failures caused by natural factors, such as aging and environmental changes, and malicious attacks driven by external human interventions \cite{artime2024robustness}. 

To address the two difficulties, we propose a method named \textbf{Meta}-heuristic \textbf{H}ypergraph-assisted \textbf{A}utomated \textbf{R}ubustness \textbf{O}ptimization \textbf{F}ramwork (Meta-HAROF), based on genetic algorithms (GAs) and hypergraph modeling. GAs, a metaheuristic algorithm from Artificial Intelligence, have been widely used to address large-scale optimization problems, due to their strong search capability \cite{miikkulainen2021biological,zhang2024survey,katoch2021review}. \textcolor{black}{Hypergraphs are effective in representing higher-order relationships among nodes \cite{antelmi2023survey,la2022music}, offering great potential to model complex higher-order systems. However, using hypergraphs for higher-order network robustness optimization remains unexplored.} 
In this study, Meta‑HAROF maps hypergraphs into bipartite graphs and introduces local and global node-hyperedge relationships rewiring operators. Building on these operators, we design novel crossover and mutation operators. The robustness of higher-order networks is evaluated using the percolation curve area \cite{schneider2011mitigation}. By incorporating adaptive integration into robustness computation, we significantly reduce computational cost while maintaining performance.

Experimental results on both synthetic and real-world hypergraphs demonstrate that Meta‑HAROF can effectively optimize network structures to enhance robustness against various types of attacks. Our analysis reveals a preferential connection pattern in optimized hypergraphs, 
where high-hyperdegree nodes tend to connect to lower-cardinality hyperedges, forming a Lotus topology. Leveraging this principle, we further propose a robust hypergraph generation method in which a single parameter, $rb$, governs the connection bias between nodes and hyperedges. Interestingly, tuning the parameter $rb$ reveals an inverse yet equally robust configuration, i.e., the Cactus topology, uncovering a previously unrecognized structural mechanism for enhancing robustness in higher-order networks. 

\section*{Measuring the Robustness of Higher-order Networks}\label{sec2}

\subsection*{Higher-order Networks Modeled by Hypergraphs}

Higher-order networks can be represented by a hypergraph. Unlike a graph, a hypergraph allows a hyperedge to connect any number of nodes. This enables a hypergraph to effectively capture complex higher-order interactions among units, extending beyond pairwise connections. Formally, a hypergraph with $N$ nodes and $M$ hyperedges can be represented by an incidence matrix $H$, where $H_{i e_\gamma} = 1$ if node $i$ is associated with hyperedge $e_\gamma$, otherwise $H_{i e_\gamma} = 0$. The hyperdegree of a node, denoted as $k_i$, corresponds to the number of hyperedges that it is associated with, while the cardinality of a hyperedge, denoted as $m_\gamma$, represents the number of nodes that it encompasses.

\begin{figure}[h]
  \begin{center}
  \includegraphics[width=5in]{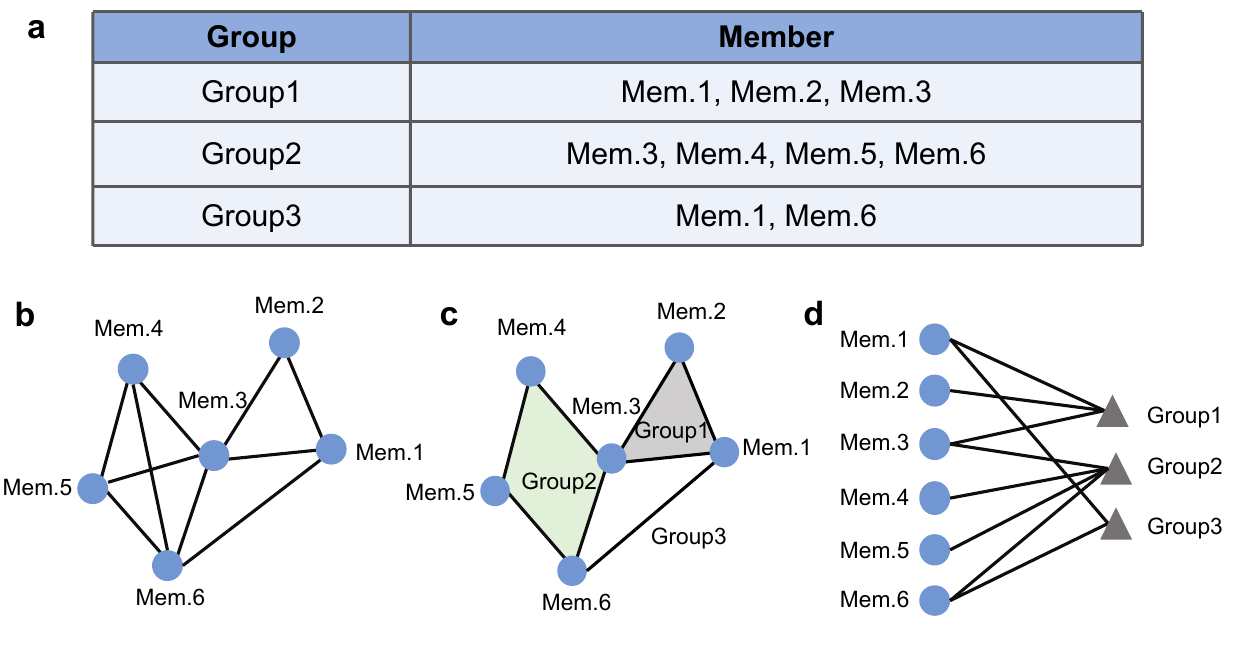}\\
  \caption{\textbf{Graph, hypergraph, and bipartite graph representations of group-member relationships.} \textbf{a}. A schematic illustration of group-member relationships in a social network. \textbf{b}. A graph representation that captures only pairwise interactions between members. \textbf{c}. A hypergraph representation that accurately models higher-order relationships by allowing a hyperedge to connect multiple members. \textbf{d}. A bipartite graph representation, providing an alternative but equivalent formulation of the hypergraph. }\label{representations}
  \end{center}
\end{figure}

Each higher-order relationship (hyperedge) can be decomposed into multiple node-hyperedge relationships, naturally leading to an equivalent bipartite representation. Specifically, a hypergraph can be interpreted as a bipartite graph $G(U, V, E)$, where $U$ represents the set of original nodes, $V$ represents the set of hyperedges, and $E$ represents the set of edges between two disjoint sets. 
In this bipartite structure, edges exist only between nodes in different subsets. The distinctions between graph, hypergraph, and bipartite representations are illustrated in Fig.~\ref{representations}.

\subsection*{Percolation Models}

Network connectivity is fundamental to maintaining its functional integrity. The percolation model \cite{morone2015influence,radicchi2015percolation} has been widely used to assess the robustness and vulnerability of complex networks under random failures or malicious attacks. Originally introduced in statistical physics \cite{Broadbent_Hammersley_1957} for describing fluid flow through porous media,  percolation theory has been successfully applied to network science to analyze the effects of node or edge failures on network structure and functionality. This inspires us to quantify the robustness of higher-order networks by evaluating the percolation curve area $R$, which characterizes how the fraction of nodes in the \textbf{L}argest \textbf{C}onnected \textbf{C}omponent (LCC) evolves under attacks \cite{schneider2011mitigation}. Accordingly, the robustness of a network subjected to node attacks $R_{N_{attack}}$, is defined as:

\begin{equation}
R_{N_{attack}} = \frac{1}{N}\sum\limits_{q = 1}^N {s(q)},
\label{Robustness-node}
\end{equation}
where $s(q)$ denotes the fraction of nodes in the LCC after removing the $q$ nodes. Similarly, we define the robustness metric for hyperedge attacks as:

\begin{equation}
R_{E_{attack}} = \frac{1}{M}\sum\limits_{q = 1}^M {s(q)}
\label{Robustness-edge}
\end{equation}
Note that $R_{N_{attack}}$ is bounded \cite{schneider2011mitigation} within $(0, 0.5]$, while $R_{E_{attack}}$ is bounded within $(0, 1]$. The upper bound of $R_{E_{attack}}$ arises in the scenario where every hyperedge connects to all the nodes in a hypergraph, resulting in maximal network connectivity. A larger $R$ indicates greater network robustness.

\section*{Evolutionary Hypergraph-assisted Automated Robustness Optimization}
\subsection*{Optimization Strategy}
For enhancing network robustness, there have been three major categories of methods, namely edge rewiring \cite{schneider2011mitigation}, redundant edge addition \cite{zhang2018optimization}, and node monitoring \cite{freitas2022graph}. In real-world systems such as power grids, edge rewiring of existing connections typically offers greater cost efficiency compared to the introduction of new network edges \cite{schneider2011mitigation}. Edge rewiring, which reconfigures connections between nodes while maintaining their degrees, has been extensively utilized in graph-based robustness optimization research \cite{huang2025enhancing, wang2023enhancing, zhou2014memetic}. Inspired by this, we extend edge rewiring to higher-order networks, where it involves reconfigures higher-order relationships while maintaining both node hyperdegree and hyperedge cardinality. In this work, we optimize $R_{N_{attack}}$ and $R_{E_{attack}}$ by rewiring higher-order relationships to enhance the robustness of higher-order networks under specific attacks.

\subsection*{Hypergraph-based Higher-order Networks Robustness Optimization}

\begin{figure}[H]
  \begin{center}
  \includegraphics[width=5.5in]{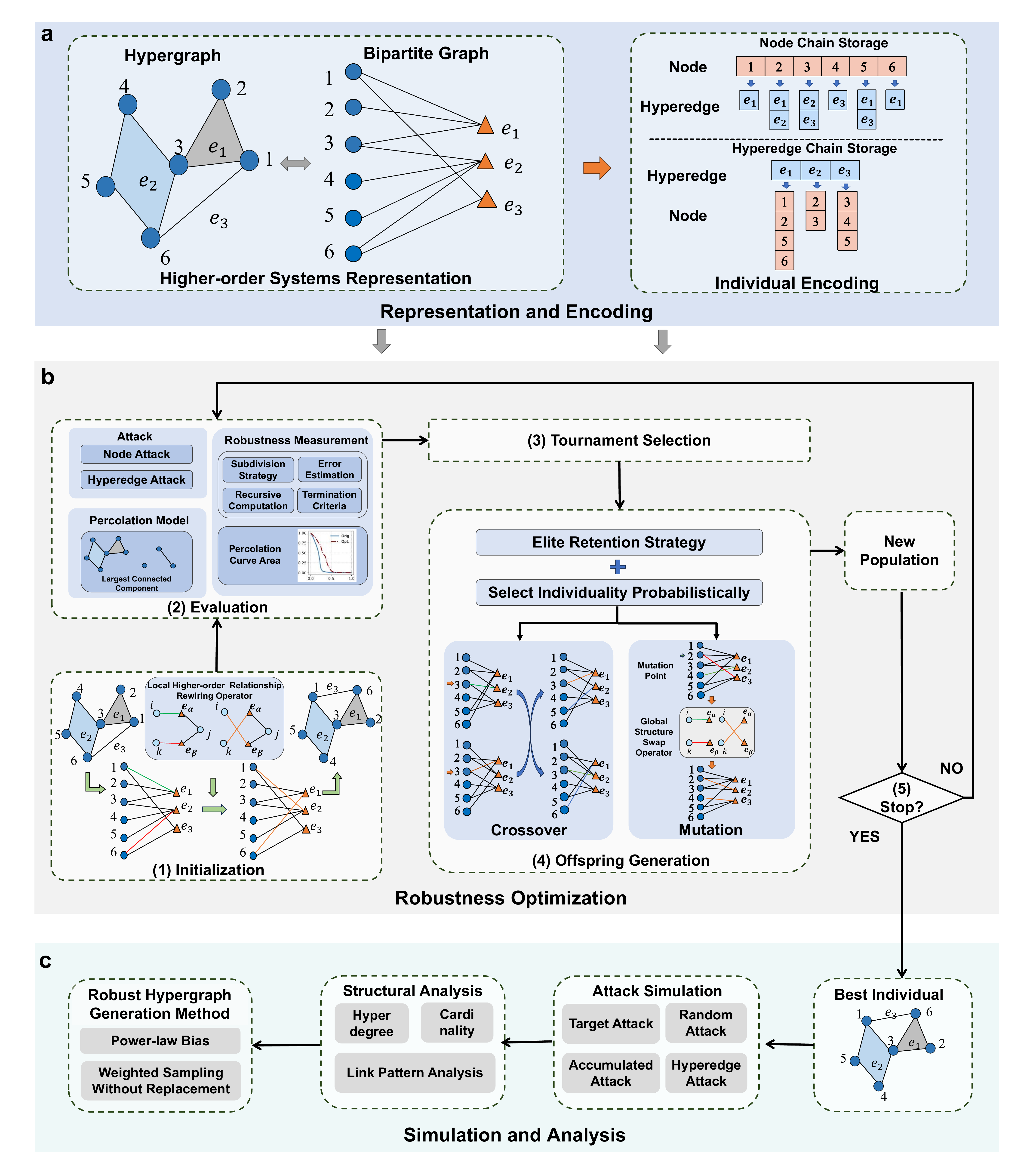}\\
  \caption{\textbf{The overall pipeline of Meta‑HAROF.} \textbf{(a)}. The Representation and Encoding module transforms the hypergraph into a bipartite graph to facilitate rewiring the higher-order interactions and employs a chain-based encoding scheme to improve computational efficiency. \textbf{(b)}. The Robustness Optimization module leverages a genetic algorithm to iteratively rewire higher-order relationships, enhancing the structural robustness. \textbf{(c)}. The Simulation and Analysis module conducts various attack simulations on the optimized hypergraph, analyzes its structural properties, and proposes a robust hypergraph generation method. }\label{framework}
  \end{center}
\end{figure}

To enhance the robustness of higher-order systems, we propose a novel Evolutionary Hypergraph-assisted Automated Robustness Optimization Framework (Meta‑HAROF). As illustrated in Fig.~\ref{framework}a, to efficiently rewire higher-order relationships, each GA individual is used to represent a hypergraph by a bipartite graph. A chain-based encoding scheme is designed to improve the computational efficiency of Meta‑HAROF. The overall framework of Meta‑HAROF is illustrated in Fig.~\ref{framework}b. The optimization process begins with initializing the population through the local higher-order rewiring operator. To accelerate the process of evaluating each individual's robustness, an adaptive integration technique is introduced. Under a specific attack, the robustness $R$ of each \textcolor{black}{individual (solution/hypergraph)} in the population is evaluated. Based on the obtained fitness values, tournament selection is employed to select better individuals to breed offspring for evolution. During evolution, an elitism-preserving strategy ensures that the most robust individuals are retained, while crossover and mutation operations are applied with certain probabilities to generate new individuals. The process iterates until a predefined termination condition is met, and outputs the best individual as the final solution.

As illustrated in Fig.~\ref{framework}c, to validate the effectiveness of Meta‑HAROF, we conduct attack simulations on the final solution (i.e., the optimized hypergraph) and perform an in-depth analysis on its connection patterns. Building upon the observed connection patterns in the optimized hypergraph, we propose a robust hypergraph generation method that enables tunable robustness against node attacks through the adjustment of a single parameter, i.e., the robustness factor.

\section*{Experimental Results of Hypergraph-assisted Robustness Optimization}

To assess the effectiveness of Meta‑HAROF, we conduct experiments on both synthetic and real-world hypergraphs. The details on experiment design, including hypergraph configurations, parameter settings, and network attacks, are provided in the \textbf{Supplementary Information Section II}.

\subsection*{Robustness Optimization of Synthetic Hypergraphs}

In the experiments on the synthetic hypergraphs, Meta‑HAROF is applied to Erdős-Rényi (ER) and Scale-Free (SF) hypergraphs \cite{aksoy2017measuring} under both malicious and hyperedge attacks, with varying parameter settings. Specifically, we considered two network scales ($N = 1,000$ and $N = 2,000$), two node-to-hyperedge ratios ($2:1$ and $1:1$), and distinct structural parameters (average hyperdegree $\left\langle k \right\rangle $ and power-law exponent $\lambda$) for the ER and SF hypergraphs. 
To ensure statistical reliability, each experiment on the synthetic hypergraphs was conducted independently using 10 hypergraphs generated with identical parameters, and this process was repeated for 30 runs. The final results were an average of 300 trials in total.

\noindent \textbf{Malicious attack:}

\begin{figure}[h]
  \begin{center}
  \includegraphics[width=5in]{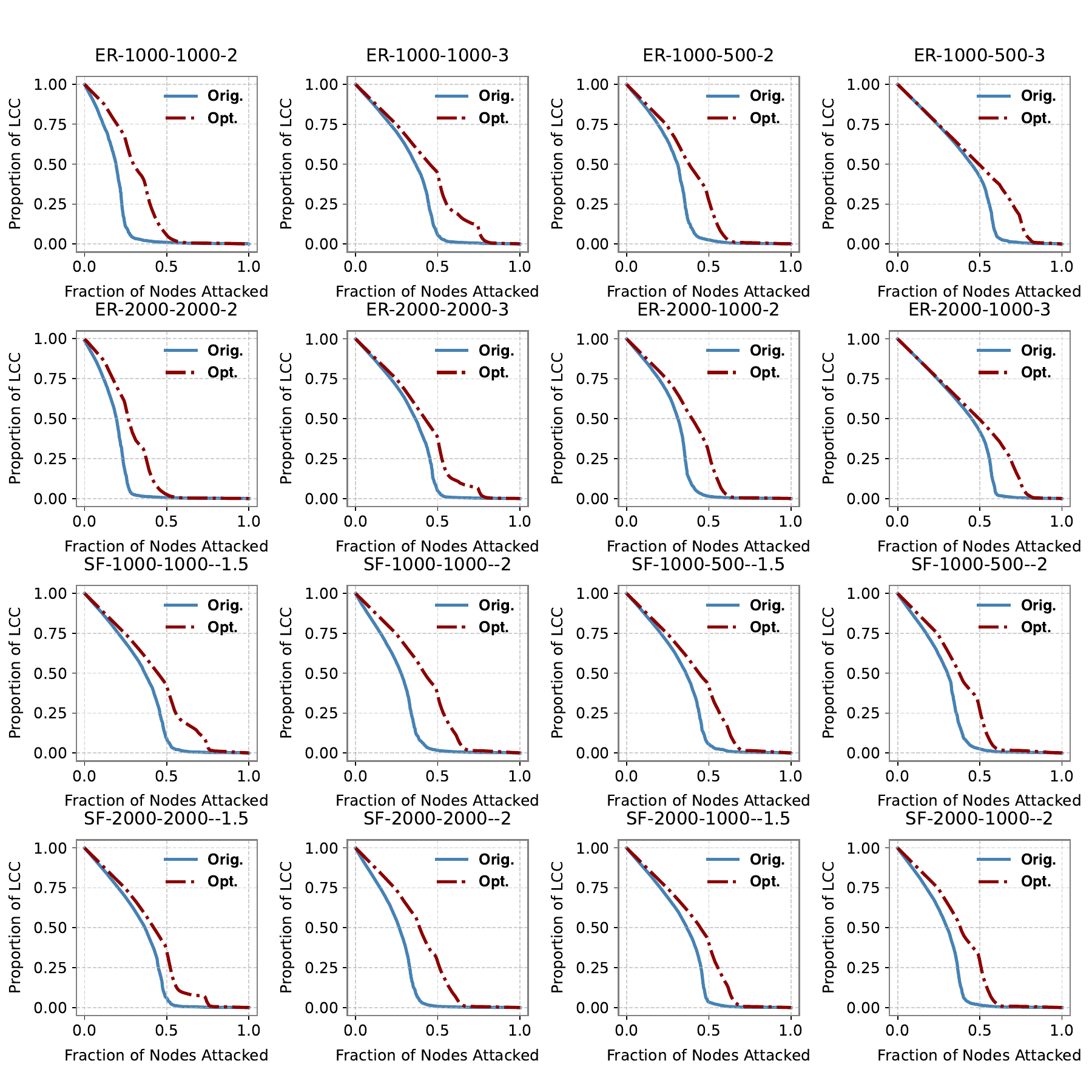}\\
  \caption{\textbf{Percolation curves of synthetic hypergraphs under malicious attack.} Each subplot illustrates the evolution of the LCC as the fraction of nodes attacked, under specific structural parameter settings, for both the original and optimized hypergraphs.
  }\label{synthetic networks target}
  \end{center}
\end{figure}

Fig. \ref{synthetic networks target} presents the percolation curves for both the original and optimized hypergraphs across various hypergraphs. 
For any given fraction of attacked nodes, the optimized hypergraphs consistently retain a larger LCC. This improvement is quantitatively captured by an increased area under the percolation curve, $R_{N_M}$. A quantitative comparison of robustness reveals that $R_{N_M} $ increases by 16.6\% to 68.7\% in the optimized hypergraphs compared to the original ones. Detailed results are provided in \textbf{Table 3 of  Section III in Supplementary Information}.

\begin{figure}[h]
  \begin{center}
  \includegraphics[width=5in]{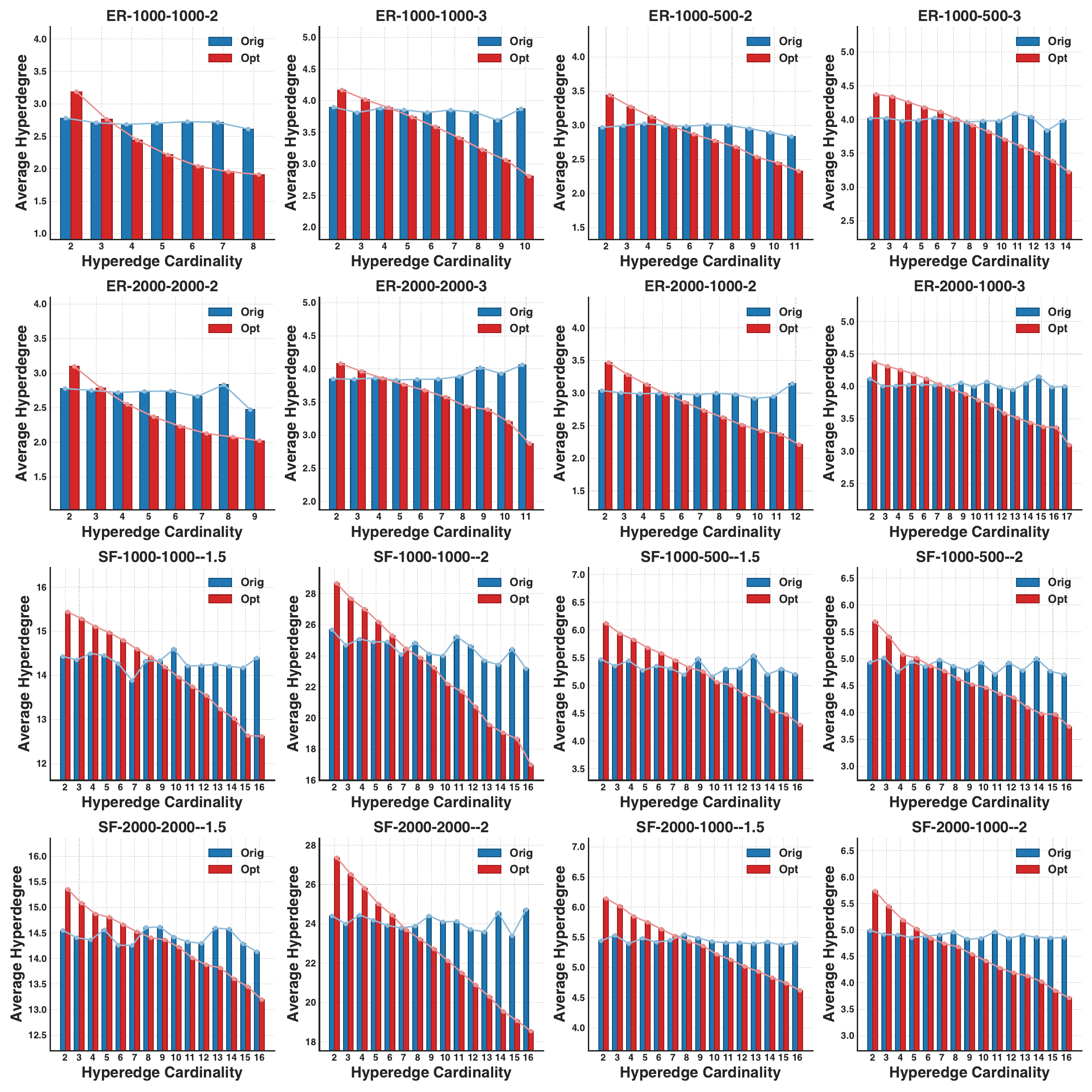}\\
  \caption{\textbf{Average hyperdegree of nodes within hyperedges of the same cardinality. The bar plot illustrates how $\langle k_m \rangle$ (y-axis) varies with $m$ (x-axis) in both original and optimized hypergraphs.} 
  }\label{bar_plot}
  \end{center}
\end{figure}

To investigate the structural changes induced by Meta‑HAROF, we analyze the node-hyperedge connection patterns in the optimized hypergraphs. Specifically, we examine whether nodes with different hyperdegrees exhibit preferential connection to hyperedges of particular cardinalities. For each hyperedge cardinality $m$, the average hyperdegree of the associated nodes, $\langle k_m \rangle$, is computed as follows:  

\begin{equation}
\left\langle k_m \right\rangle = \frac{
  \sum\limits_{\substack{e_\gamma \in E \\ m_\gamma = m}} \sum\limits_{i \in e_\gamma} k_i
}{
  \sum\limits_{\substack{e_\gamma \in E \\ m_\gamma = m}} m_\gamma
}.
\label{ave}
\end{equation}

The results, visualized in Fig. \ref{bar_plot}, indicate the relationship between hyperedge cardinality (x-axis) and the average hyperdegree of the associated nodes (y-axis). In the original hypergraph, no significant correlation is observed between these two quantities. However, across all the optimized hypergraphs, a clear inverse relationship emerges, i.e., \textbf{nodes with higher hyperdegrees tend to connect lower-cardinality hyperedges}. This structural configuration gives rise to the distinctive \textbf{Lotus topology}, where high-hyperdegree nodes exhibit selective connection to lower-cardinality hyperedges, as illustrated in Fig. \ref{realnetworks target}b. If this connection pattern is a general feature of robust hypergraphs, it would offer valuable principles for designing robust higher-order networks.

\noindent \textbf{Hyperedge attack:}

To further examine the generalizability of Meta‑HAROF, we apply it to enhance hypergraph robustness against hyperedge attacks through optimization of Eq. (\ref{Robustness-edge}). \textbf{Supplementary Information Section IV} presents the percolation curves and the comparison of robustness $R_{M_H}$ between the original and optimized hypergraphs. The results demonstrate that the optimized hypergraphs consistently exhibit a larger percolation curve area and higher robustness, with $R_{E_{attack}}$ improving by 18.2\% to 45.6\% compared to the original hypergraphs. These findings confirm the generalizability of Meta‑HAROF in enhancing robustness against both the node and hyperedge attacks.

\subsection*{Robustness Optimization for Real-world Hypergraphs}

\begin{figure}[h]
  \begin{center}
  \includegraphics[width=5.2in]{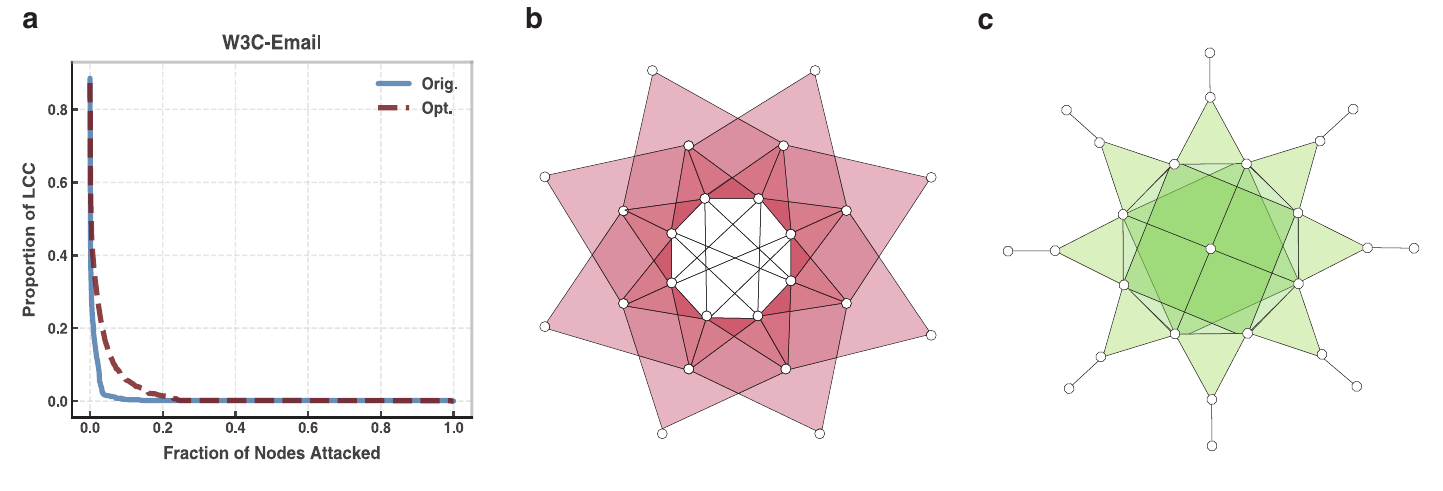}\\
  \caption{\textbf{Percolation curves of W3C-Email hypergraphs and schematic diagram of Lotus and Cactus topology.} \textbf{(a)}. Percolation curves of the W3C-Email hypergraph under malicious attack.   \textbf{(b)}. Schematic representations of Lotus topology.  \textbf{(c)}. Schematic representations of Cactus topology.
  }\label{realnetworks target}
  \end{center}
\end{figure}

To evaluate the real-world effectiveness of the Meta‑HAROF, we analyze its performance on a communication network constructed from emails exchanged on W3C mailing lists \cite{amburg2021planted,craswell2005overview}. The W3C-Email hypergraph consists of 14,317 nodes and 19,821 hyperedges, where nodes represent individual email addresses and hyperedges correspond to groups of addresses appearing together in the same email. Due to memory constraints, the population size was set to 128 in the real-world hypergraph experiments, while all other settings were kept consistent with those used in the synthetic hypergraph experiments. The percolation curves of the original and optimized hypergraphs are shown in Fig. \ref{realnetworks target}a. The results indicate that under the same proportion of attacked nodes, the optimized hypergraph retains a higher fraction of nodes in the LCC, particularly in the early stages of the attack, and exhibits a larger percolation curve area. The robustness metric $R_{N_M}$ against malicious attacks improves from 0.0077 to 0.0235, an increase of over 200\%. This experiment demonstrates the potential applicability of Meta‑HAROF in communication networks and similar real-world scenarios.

\section*{Robust Hypergraph Generation Method}

\begin{figure}[h]
  \begin{center}
  \includegraphics[width=5in]{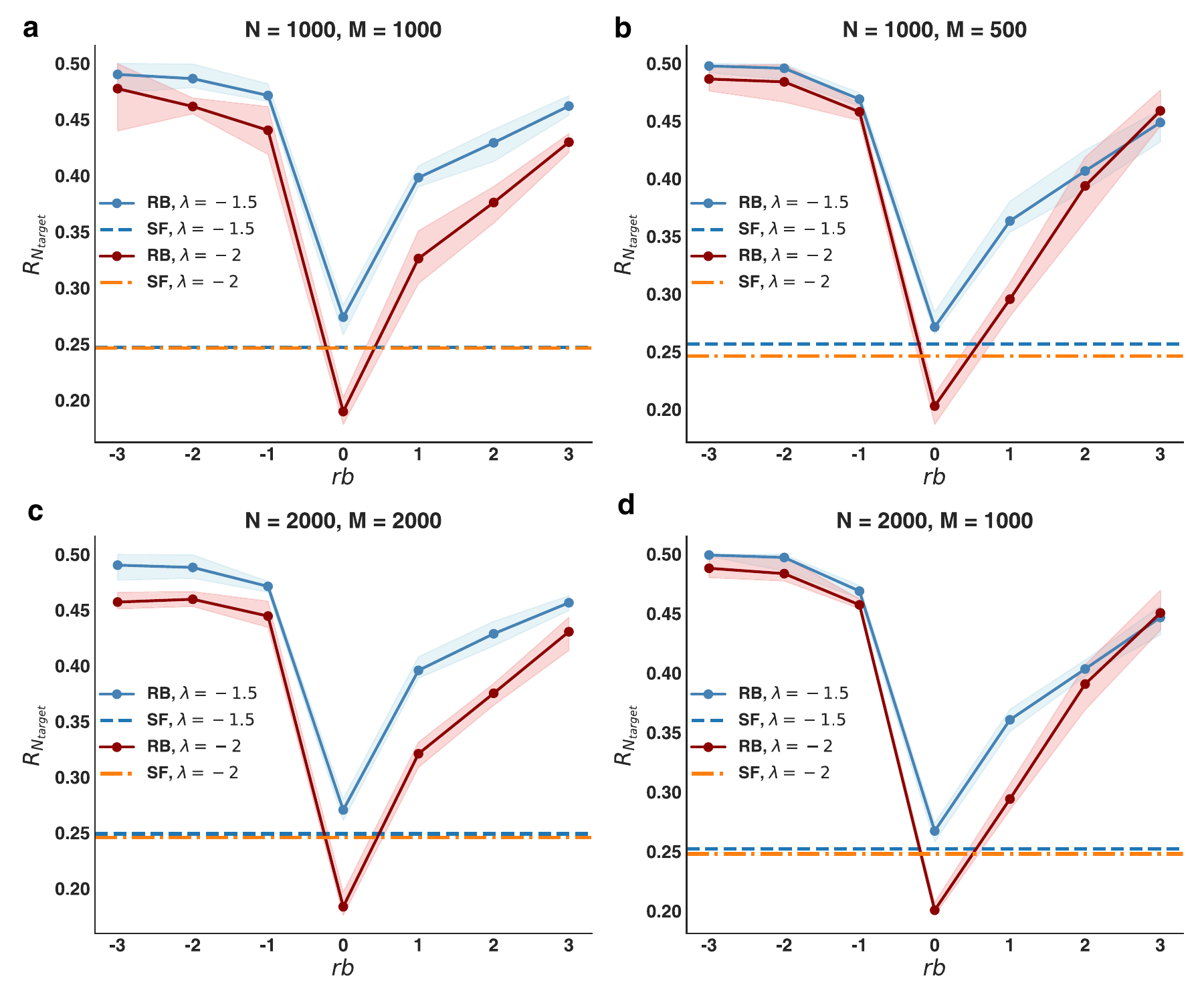}\\
  \caption{\textbf{Robustness of RB and SF hypergraphs under malicious attack.} Robustness of RB and SF hypergraphs is evaluated under malicious attack, with structural parameters held constant except for varying $rb$ values in RB hypergraphs. Shaded regions represent the variability in RB robustness across different realizations.
  }\label{target_robustness}
  \end{center}
\end{figure}

To validate the empirical observation that nodes with higher hyperdegrees tend to connect lower-cardinality hyperedges in optimized hypergraphs, we propose \textbf{a Robust (RB) Hypergraph Generation Method} that incorporates Power-law Bias and Weighted Sampling Without Replacement. The robust hypergraph generation method begins with the hyperedge cardinality sequence $\{m_\gamma\}$ and node hyperdegree sequence $\{k_i\}$. A global allocation table $remaining[i] = k_i$ is initialized, and hyperedges are sorted in ascending cardinality to prioritize lower-cardinality hyperedges. Nodes are assigned via weighted sampling without replacement, where the selection weight follows a power-law bias:

\begin{equation}
\begin{array}{l}
{w_i} = k_i^{{\beta _\gamma }},\quad \\
{\beta _\gamma } = \frac{{\max (\{ {m_\gamma }\} )}}{{{m_\gamma }}} \cdot rb,
\end{array}
\end{equation}
where $rb$ is a tunable robustness parameter that controls the extent of preferential connection. The term $\frac{\max(\{m_\gamma\})}{m_\gamma}$ ensures that lower-cardinality hyperedges correspond to larger values of $\beta_\gamma$. When $rb>1$, lower-cardinality hyperedges have larger $\beta_\gamma$, ensuring higher-hyperdegree nodes preferentially connect to them. When $rb<-1$, lower-cardinality hyperedges have smaller $\beta_\gamma$, ensuring lower-hyperdegree nodes preferentially connect to them. The sampling process proceeds iteratively, progressively reducing node quotas until all hyperedges are populated while maintaining hyperdegree constraints. Further details are provided in \textbf{Section V A in Supplementary Information.}


We utilize the RB hypergraph generation method to construct robust power-law-distributed hypergraphs with varying robustness controlled by the parameter $rb$, and subsequently compare their robustness with the SF hypergraphs generated under identical structural parameters, including power-law distribution, size, and node-to-hyperedge ratio. As shown in Fig. \ref{target_robustness}, the robustness of RB hypergraphs exhibits distinct trends with varying $rb$ values. When $rb > 1$, higher-hyperdegree nodes tend to connect to the hyperedges of lower-cardinality, and this tendency becomes more pronounced as $rb$ increases, resulting in a \textbf{Lotus topology} that enhances robustness. Conversely, when $rb < -1$, lower-hyperdegree nodes are more likely to connect to lower-cardinality hyperedges, and robustness improves as $rb$ decreases, forming a novel robust \textbf{Cactus topology}, as illustrated in Fig. \ref{realnetworks target}c. When $rb = 0$, the connection between nodes and hyperedges shows no preferential bias. The results of the experiment show that both the Lotus and Cactus topologies exhibit superior robustness, with both connection preferences and overall robustness being tuned via a single control parameter $rb$. \textcolor{black}{The enhanced robustness of lotus-topology and cactus-topology hypergraphs against malicious attacks may stem from the tightly connected critical nodes, preventing structural disintegration.} Moreover, it is found that when the lotus-topology hypergraph converts to a graph by restricting all the hyperedges to cardinality two, the forming topology naturally becomes an onion-like structure \cite{schneider2011mitigation}, a well-known robust structure in conventional graphs. This observation highlights the universality and extensibility of the proposed structural paradigm beyond hypergraph settings. 

\section*{Discussions}



In this work, we proposed Meta‑HAROF, a novel framework for enhancing the robustness of higher-order networks by decomposing higher-order relationships into node-hyperedge relationships and strategically rewiring them. This approach addresses a critical gap in hypergraph-based robustness optimization and demonstrates its effectiveness through extensive experiments.

A key insight emerging from our analysis is the preferential connection pattern in optimized hypergraphs: nodes with higher hyperdegree tend to connect with hyperedges of lower cardinality. This leads to a distinctive Lotus topology. The structural connection tendency significantly enhances robustness against malicious attacks, underscoring the potential of Meta‑HAROF for designing robust higher-order networks. Inspired by this observation, we have further proposed a robust hypergraph generation method, enabling robustness to be \textcolor{black}{controlled} by a single parameter, $rb$. Notably, for $rb < -1$, a distinct Cactus topology emerges, contrasting with the Lotus topology observed for $rb > 1$. Despite their structural differences, both topologies exhibit strong robustness to malicious attacks. The discovery of the Lotus and Cactus topologies offers valuable insights for designing robust higher-order networks, while also providing important implications for the analysis of higher-order network structures and cascading dynamics. Although Meta‑HAROF achieves enhanced performance through adaptive integration, its computational demands impose scalability limitations for large-scale network applications. Future efforts should focus on developing more efficient optimization strategies by incorporating domain-specific priors. 

\bibliography{sn-bibliography}

\section*{Methods}
This section elaborates on the crucial components of Meta‑HAROF. It mainly consists of the encoding strategy, initialization, individual evaluation, crossover, and mutation operators. We first introduce the local and global higher-order relationship rewiring operators, which serve as the theoretical foundation for the initialization method and mutation operator in Meta‑HAROF. 

\subsection*{Local Higher-order Relationship Rewiring Operator}

The local higher-order relationship rewiring operator primarily adjusts higher-order relationships by locally modifying the node-hyperedge relationship in a bipartite graph while preserving node hyperdegree and hyperedge cardinality. The definition of the local higher-order relationship rewiring operator $\mathcal{L}$ is as follows:
\begin{equation}
{\cal L}({H_{i{e_\gamma }}},{H_{j{e_\gamma }}},{H_{j{e_\beta }}},{H_{k{e_\beta }}}) = ({H_{i{e_\beta }}},{H_{j{e_\gamma }}},{H_{j{e_\beta }}},{H_{k{e_\gamma }}}),
\label{local}
\end{equation}
where $H_{*e_{*}}$, the first subscript indicates node $*$ (including $i$, $j$, or $k$), and the second subscript indicates the hyperedge $e_{*}$ (including $e_\beta$ or $e_\gamma$). Note that $i\in e_\gamma$, $j \in e_\gamma/i$, $e_\beta \in \partial j/{e_\gamma }$, $k \in e_\beta/j$, and $k \ne i$. Specifically, $e_\gamma/i$ represents the set of nodes in hyperedge $e_\gamma$ excluding node $i$, and $\partial j/{e_\gamma }$ represents the set of hyperedges related to node $j$ excluding hyperedge $e_\gamma$. More details about the local higher-order relationship rewiring operator are illustrated in \textbf{Supplementary Information Section II A}.

\subsection*{Global Higher-order Relationship Rewiring Operator}

The global higher-order relationship rewiring operator primarily adjusts 
higher-order relationships through globally modifying the node-hyperedge relationship in a bipartite graph while preserving node hyperdegree and hyperedge cardinality. For any two node-hyperedge relationship $H_{ie_\gamma}$ and $H_{ke_\beta}$, where $i \neq k$ and $e_\gamma \neq e_\beta$, the definition of the global higher-order relationship rewiring operator $\mathcal{G}$ is as follows:
\begin{equation}
{\mathcal{G}}({H_{i{e_\gamma }}},{H_{k{e_\beta }}}) = ({H_{i{e_\beta }}},{H_{k{e_\gamma }}})
\label{global}
\end{equation}
Note that the global higher-order relationship rewiring operator can work without constraints from local neighboring node-hyperedge relationships, facilitating arbitrary exchange between any two non-neighbor node-hyperedge relationships across the entire bipartite graph. The details of the global higher-order relationship rewiring operator are introduced in \textbf{Supplementary Information Section II B}.

\subsection*{Hypergraph Encoding}

In the proposed Meta‑HAROF framework, each individual is used to represent a hypergraph. To enhance computational efficiency and reduce memory consumption, a dual-chain representation is designed to encode a hypergraph. This design significantly reduces the time complexity of neighborhood access operations from $\mathcal{O}(N)$ to $\mathcal{O}(1)$. 
Specifically, the individual encoding scheme comprises two components: Node Chain Storage and Hyperedge Chain Storage. In the Node Chain Storage, the index of the array corresponds to a node $id$, and each node links to a chain that stores the $id$ of its incident hyperedges. Conversely, in the Hyperedge Chain Storage, the index corresponds to a hyperedge $id$, and each hyperedge connects to a chain that contains the $id$ of its constituent nodes. This bidirectional mapping not only reduces the memory footprint of the hypergraph representation but also enables constant-time access to neighboring nodes and hyperedges, thereby facilitating effective evolutionary operations in hypergraphs.

\subsection*{Initialization}

The initialization is designed based on the local higher-order relationship rewiring operator. Given the original hypergraphs $H^0$, we randomly select $r^* |H^0|$ node-hyperedge relationships from $H^0$ to perform the local higher-order relationship rewiring operator, thereby generating an individual. Here, $|H^0|$ represents the number of node-hyperedge relationships, and $r$ is a random number uniformly distributed between $0$ and $1$. The original and newly generated higher-order networks collectively constitute the population. The illustration of the initialization operator is shown in \textbf{Supplementary Information Section II C}.

\subsection*{Adaptive Integration For Individual Evaluation}

To enhance the computational efficiency of robustness evaluation in hypergraphs, adaptive integration \cite{davis1984methods} is employed to dynamically adjust the sampling density based on percolation curve variability. 
Instead of uniform summation in Eq.(\ref{Robustness-node}) and Eq.(\ref{Robustness-edge}), the adaptive Simpson’s rule is used to allocate more evaluation points near critical transitions, therefore improving accuracy with fewer function evaluations and reducing computational overhead. Statistical analysis of the computation time for network robustness $R$ demonstrates that adaptive integration can improve efficiency by a factor of \textbf{3.3× to 43.8×}. More details can be found in \textbf{Supplementary Information Section I}.

\subsection*{Crossover Operator}

We design a novel uniform crossover operator based on the bipartite graph representation, leveraging the structural characteristics of hypergraphs. In this representation, each node in the bipartite graph has a certain probability $p_c$ to be selected as a crossover point. 
If node $i$ is selected as a crossover point, we then examine whether node $i$ in bipartite graphs $B^p$ and $B^q$ is associated with different hyperedges. Specifically, if node $i$ is associated with hyperedge $e_\alpha$ in $B^p$ and with hyperedge $e_\beta$ in $B^q$, we swap the node-hyperedge relationships $H_{ie_\alpha}$ and $H_{ie_\beta}$. The specific steps are as follows:

\textbf{For $B^p$:} we first identify a node $j$ that is associated with hyperedge $e_\beta$ but not hyperedge $e_\alpha$. Afterwards, the swap operator $\mathcal{S}$ to exchange node-hyperedge relationship is defined as follows:

\begin{equation}
{\cal S}({H_{i{e_\alpha }}},{H_{j{e_\beta }}}) = ({H_{i{e_\beta }}},{H_{j{e_\alpha }}})
\label{cross1}
\end{equation}

\textbf{For $B^q$:} we first identify a node $k$ that is associated with hyperedge $e_\alpha$ but not hyperedge $e_\beta$. Then, the node-hyperedge relationships are swapped by:

\begin{equation}
{\cal S}({H_{k{e_\alpha }}},{H_{i{e_\beta }}}) = ({H_{k{e_\beta }}},{H_{i{e_\alpha }}})
\label{cross2}
\end{equation}
These operations enable the exchange of node-hyperedge relationship $H_{ie_\alpha}$ in $B^p$ with $H_{ie_\beta}$ in $B^q$. The crossover operator is illustrated in \textbf{Supplementary Information Section II D}.

\subsection*{Mutation Operator}

To improve Meta‑HAROF's search capability and prevent it from getting trapped in local optima, we design a multi-point mutation operator that prioritizes the mutation of key nodes for maintaining system robustness \cite{qu2025efficient}. Since the hyperdegree of a node serves as a fundamental centrality measure that effectively captures its importance \cite{xie2023efficient}, we incorporate it into the mutation process to guide node selection. Specifically, the hyperdegree of each node is scaled by a random factor within the range of [0.5, 1], after which the nodes are sorted in descending order. The top $r_{m}$ nodes are then selected as mutation points, where $r_m$ is a hyperparameter that controls the proportion of mutated nodes. For a selected mutation point $i$, a node-hyperedge relationship $H_{ie_\alpha}$ associated with node $i$ is randomly chosen. Subsequently, another node-hyperedge relationship $H_{je_\beta}$ is randomly selected, ensuring that $j \notin e_\alpha$ and $i \notin e_\beta$. Finally, the global higher-order relationship rewiring operator is applied to reconfigure the selected relationships $H_{ie_\alpha}$ and $H_{je_\beta}$. The mutation operator is illustrated in \textbf{Supplementary Information Section II E}.

\backmatter

\bmhead{Supplementary Materials}

The supplementary materials are included in the file of Supplementary Information.

\bmhead{Acknowledgements}

This work was supported in part by the National Key Research and Development Program of China under grant 2024YFA1012700, the National Natural Science Foundation of China under grants 62206041, 12371516 and U21A20491, and the NSFC-Liaoning Province United Foundation under grant U1908214, the 111 Project under grant D23006, the Liaoning Revitalization Talents Program under grant XLYC2008017, and China University Industry-University-Research Innovation Fund under grants 2022IT174, Natural Science Foundation of Liaoning Province under grant 2023-BSBA-030, and an Open Fund of National Engineering Laboratory for Big Data System Computing Technology under grant SZU-BDSC-OF2024-09. 

\section*{Declarations}

\subsection*{Author contribution}

\subsection*{Conflict of interest}
The authors declare no competing interests.







\begin{appendices}




\end{appendices}


\end{document}


\title[Article Title]{Supplementary Information for Meta-heuristic Hypergraph-Assisted Robustness Optimization for Higher-order Complex Systems}

\maketitle

\tableofcontents

\section{Adaptive Integration}

To enhance the computational efficiency of robustness evaluation in hypergraphs, we employ adaptive integration to compute the robustness metrics $R_{N_{attack}}$ and $R_{M_{attack}}$. Traditional uniform summation struggles to efficiently capture critical transitions in percolation processes, leading to unnecessary function evaluations. By integrating adaptive Simpson’s rule, we refine the selection of evaluation points, allocating higher resolution near abrupt changes in the percolation curve while maintaining computational efficiency. The detailed procedure of adaptive integration and its computational time analysis are presented as follows.

\subsection{The Procedure of the Adaptive Integration}

A bisection-based subdivision strategy is employed to dynamically adjust the evaluation intervals. Initially, the robustness function is evaluated at three key points: the start ($a$), midpoint ($m$), and end ($b$) of the interval. The integrand, $s(q)$, which denotes the fraction of nodes in the LCC after removing the $q$ nodes, is first evaluated at three key points: the endpoints $a$ and $b$, and the midpoint $m = \frac{a + b}{2}$. An initial approximation of the integral is obtained based on Simpson’s rule as follows:

\begin{equation}
S(a, b) = \frac{(b - a)}{6} \left[ s(a) + 4s\left( \frac{a + b}{2} \right) + s(b) \right].
\end{equation}

To assess the local accuracy of this estimate, the interval $[a, b]$ is subdivided into two subintervals: $[a, m]$ and $[m, b]$. Each subinterval is then evaluated by Simpson’s rule with newly introduced midpoint evaluations at $\frac{a + m}{2}$ and $\frac{m + b}{2}$, respectively. The refined integral estimate is given by:

\begin{equation}
\hat{S} = S_{\text{left}} + S_{\text{right}},
\end{equation}
where

\begin{align}
S_{\text{left}} &= \frac{(m - a)}{6} \left[ s(a) + 4s\left( \frac{a + m}{2} \right) + s(m) \right], \\
S_{\text{right}} &= \frac{(b - m)}{6} \left[ s(m) + 4s\left( \frac{m + b}{2} \right) + s(b) \right].
\end{align}

The difference between the two estimates, $|\hat{S} - S(a, b)|$, can be used as an estimate of the local integration error. If this error exceeds a predefined tolerance $\varepsilon$, the interval is recursively subdivided and the integration process is repeated on each subinterval. This recursive refinement selectively increases the sampling resolution in regions where the integrand exhibits rapid variations—such as near phase transitions or percolation thresholds—while maintaining a coarser resolution in smoother regions. This adaptivity significantly reduces the number of required function evaluations without compromising overall accuracy. Considering that the optimized network consists of several thousand nodes, we set the maximum recursion depth $d_{\max} = 10$ and error threshold $\varepsilon = 10^{-4}$.

\subsection{Computation Time}

To assess the computational efficiency and precision of the adaptive integration method, we conducted experiments on hypergraphs with varying structural parameters to compare its execution time with that of the original integration method. Besides, we quantified the average error between the two methods to evaluate the precision of the adaptive integration method. The hypergraphs follow the naming convention ER-$N$-$M$-$k$ for ER hypergraphs and SF-$N$-$M$-$\gamma$ for SF hypergraphs, where $N$ denotes the number of nodes; $M$ denotes the number of hyperedges; $k$ and $\gamma$ represent the average hyperdegree and power-law exponent, respectively. To ensure the reliability of results, each experiment trial has been independently conducted ten times, with the reported values representing the average computation time and error rate across the total trials. All the experiments have been conducted on an Intel Ultra 5 125H processor. The results, summarized in Table \ref{tab:runtime_comparison}, indicate that the adaptive integration method achieves a speedup ranging from 3.3× to 43.8× compared to the original integration method. Meanwhile, the maximum observed error rate remains as low as 0.004, indicating that the proposed method significantly enhances computational efficiency while preserving high precision in evaluating the robustness metric $R_{N_{attack}}$.

\begin{table}[h]
    \centering
    \caption{Comparison of Execution Time and Error between Original Integration (Orig.) and Adaptive Integration (Adapt.) Methods}
    \begin{tabular}{l S[table-format=3.6] S[table-format=3.6] S[table-format=1.6]}
        \toprule
        \textbf{Network} & \textbf{Orig. Time (s)} & \textbf{Adapt. Time (s)} & \textbf{Err.} \\
        \midrule
        ER-1000-500-2   & 0.451725  & \textbf{0.136015}  & 0.000017 \\
        ER-1000-500-3   & 0.642731  & \textbf{0.054696}  & 0.000209 \\
        ER-1000-1000-2  & 0.306340  & \textbf{0.052684}  & 0.000195 \\
        ER-1000-1000-3  & 0.524791  & \textbf{0.074422}  & 0.000162 \\
        ER-2000-1000-2  & 1.618356  & \textbf{0.127952}  & 0.000143 \\
        ER-2000-1000-3  & 2.992700  & \textbf{0.172931}  & 0.000007 \\
        ER-2000-2000-2  & 1.322420  & \textbf{0.098477}  & 0.000062 \\
        ER-2000-2000-3  & 2.185139  & \textbf{0.191605}  & 0.000295 \\
        SF-1000-500--1.5  & 0.456107  & \textbf{0.004897}  & 0.004008 \\
        SF-1000-500--2    & 0.342173  & \textbf{0.060026}  & 0.000223 \\
        SF-1000-1000--1.5 & 0.596369  & \textbf{0.080418}  & 0.000021 \\
        SF-1000-1000--2   & 0.406487  & \textbf{0.037735}  & 0.000498 \\
        SF-2000-1000--1.5 & 1.977350  & \textbf{0.132338}  & 0.000055 \\
        SF-2000-1000--2   & 1.389226  & \textbf{0.175209}  & 0.000076 \\
        SF-2000-2000--1.5 & 2.588648  & \textbf{0.191816}  & 0.000247 \\
        SF-2000-2000--2   & 1.709420  & \textbf{0.044715}  & 0.001390 \\
        Email-W3C         & 90.536543 & \textbf{2.065322} & 0.000101 \\
        \bottomrule
    \end{tabular}
    \label{tab:runtime_comparison}
\end{table}

\section{Major Components in the Meta‑HAROF Framework}


The Meta‑HAROF framework enhances the robustness of higher-order networks through higher-order relationship rewiring. To ensure that both node hyperdegree and hyperedge cardinality remain unchanged, we adopt a bipartite graph representation and design components in Meta‑HAROF based on node-hyperedge relationships. The details of these components are described as follows.

\subsection{Local Higher-order Relationship Rewiring Operator}

The local higher-order relationship rewiring operator primarily adjusts higher-order relationships by locally modifying the node-hyperedge relationship. 
For example, in a hypergraph, node $i$ is initially associated with hyperedge $e_{\gamma}$, forming the relationship $H_{i e_{\gamma}}$. Similarly, node $j$ is associated with both $e_{\gamma}$ and another hyperedge $e_{\beta}$ (i.e., forming the relationships $H_{j e_{\gamma}}$ and $H_{j e_{\beta}}$), and node $k$ is associated with $e_{\beta}$ (i.e., forming the relationship $H_{k e_{\beta}}$). 
The initial node-hyperedge relationships are illustrated in Fig.~\ref{Local-operator} a. By applying the local higher-order relationship rewiring operator $\mathcal{L}$, the node-hyperedge relationships are adjusted as follows:  

\begin{enumerate}

\item Node $i$ is reassigned from hyperedge $e_{\gamma}$ to $e_{\beta}$, forming $H_{i e_{\beta}}$ instead of $H_{i e_{\gamma}}$. 
\item Node $k$ is reassigned from hyperedge $e_{\beta}$ to $e_{\gamma}$, forming $H_{k e_{\gamma}}$ instead of $H_{k e_{\beta}}$.
\item The connections of node $j$ remain unchanged, preserving the relationships $H_{j e_{\gamma}}$ and $H_{j e_{\beta}}$. 

\end{enumerate} 
  
After applying the operator, the updated node-hyperedge relationships are shown in Fig.~\ref{Local-operator} b. This transformation preserves both the node hyperdegree and hyperedge cardinality while locally adjusting the higher-order connectivity structure.
\renewcommand{\figurename}{Fig.} 
\renewcommand{\thefigure}{S\arabic{figure}} 
\begin{figure}[h]
  \begin{center}
  \includegraphics[width=2.5in]{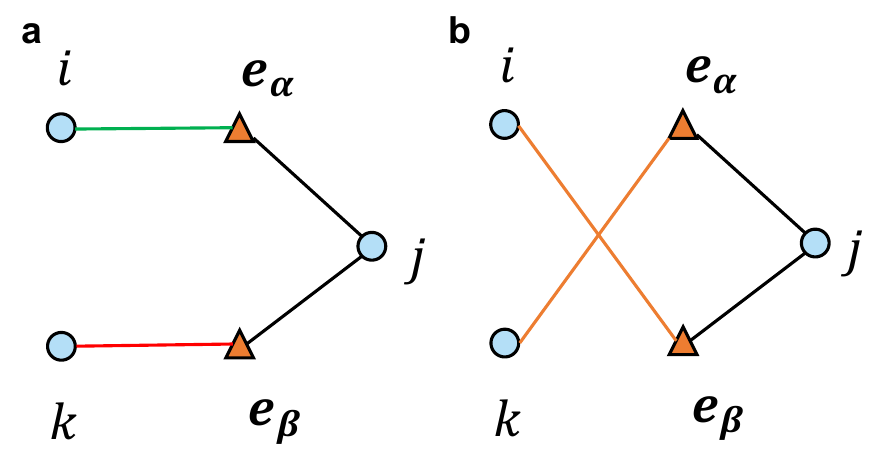}\\
  \caption{The illustration of the local higher-order relationship rewiring operator.}\label{Local-operator}
  \end{center}
\end{figure}

\subsection{Global Higher-order Relationship Rewiring Operator}

\begin{figure}[h]
  \begin{center}
  \includegraphics[width=2in]{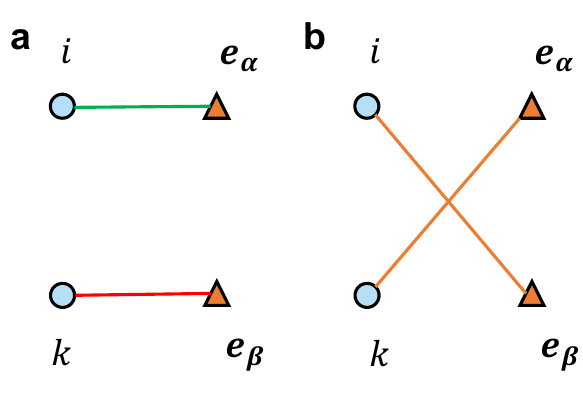}\\
  \caption{The illustration of the global higher-order relationship rewiring operator.}\label{Global-operator}
  \end{center}
\end{figure}

The global higher-order relationship rewiring operator primarily adjusts higher-order relationships by globally modifying the node-hyperedge relationships. For example, in a hypergraph, node $i$ is initially associated with hyperedge $e_{\gamma}$, forming the relationship $H_{i e_{\gamma}}$; node $k$ is associated with another hyperedge $e_{\beta}$, forming the relationship $H_{k e_{\beta}}$. The initial node-hyperedge relationships are illustrated in Fig.~\ref{Global-operator} a.  

By applying the global higher-order relationship rewiring operator $\mathcal{G}$, the node-hyperedge relationships are adjusted as follows:

\begin{enumerate}
\item  Node $i$ is reassigned from hyperedge $e_{\gamma}$ to $e_{\beta}$, forming $H_{i e_{\beta}}$ instead of $H_{i e_{\gamma}}$.  
\item  Node $k$ is reassigned from hyperedge $e_{\beta}$ to $e_{\gamma}$, forming $H_{k e_{\gamma}}$ instead of $H_{k e_{\beta}}$.  
\end{enumerate}

Unlike the local operator, the global higher-order relationship rewiring operator is not constrained by neighboring node-hyperedge relationships and can facilitate the exchange of any two non-neighbor node-hyperedge relationships. After applying the operator, the updated node-hyperedge relationships are shown in Fig.~\ref{Global-operator} b. This demonstrates how this transformation enables a more flexible and extensive reconfiguration of the higher-order relationships.

\subsection{Initialization}

\begin{figure}[h]
  \begin{center}
  \includegraphics[width=5in]{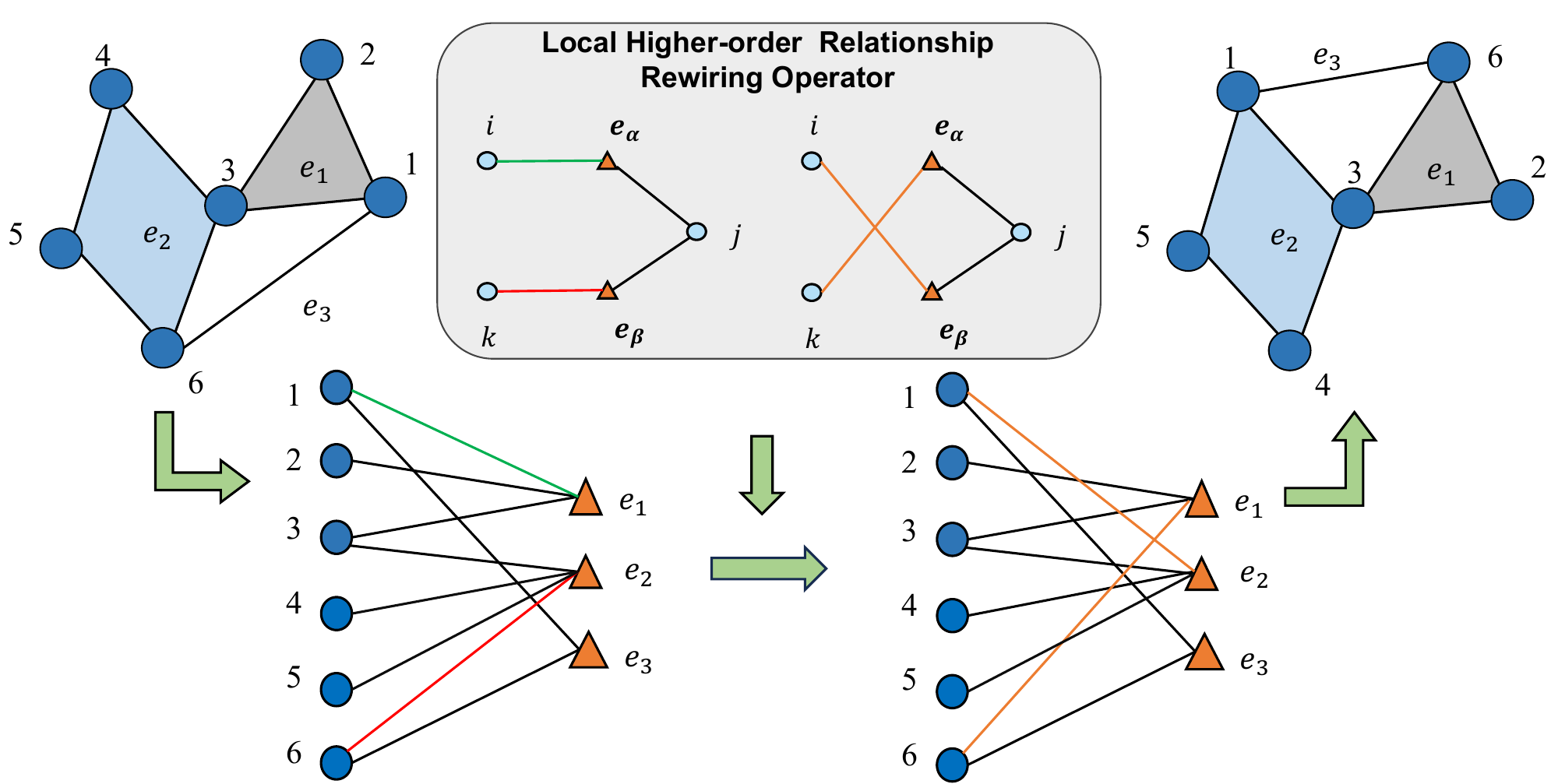}\\
  \caption{The illustration of the initialization step.}\label{initialization-operator}
  \end{center}
\end{figure}

The initialization step is designed based on the local higher-order relationship rewiring operator to generate diverse initial higher-order network structures. For example, considering an initial hypergraph $H^0$ with six nodes and three hyperedges:

\begin{itemize}
\item Hyperedge $e_1$ connects nodes $\{ 1, 2, 3 \}$, forming the node-hyperedge relationships $H_{1 e_1}, H_{2 e_1}$, and $H_{3 e_1}$.  
\item Hyperedge $e_2$ connects nodes $\{ 3, 4, 5, 6 \}$, forming the relationships $H_{3 e_2}$, $H_{4 e_2}$, $H_{5 e_2}$, and $H_{6 e_2}$.  
\item Hyperedge $e_3$ connects nodes $\{ 1, 6 \}$, forming the relationships $H_{1 e_3}$ and $H_{6 e_3}$.
\end{itemize}

The initial hypergraph is illustrated on the left side of Fig.~\ref{initialization-operator}. To create a new individual, $r^* |H^0|$ node-hyperedge relationships are randomly selected from $H^0$, subsequently applying the local higher-order relationship rewiring operator. Supposing that the relationship $H_{c e_2}$ is selected for rewiring, we then identify the neighboring node-hyperedge relationship $H_{e e_3}$. By applying the operator, the node-hyperedge relationships are adjusted as follows:  

\begin{itemize}
\item Node $1$ is reassigned from hyperedge $e_1$ to $e_2$, forming $H_{1 e_2}$ instead of $H_{1 e_1}$.  
\item Node $6$ is reassigned from hyperedge $e_2$ to $e_1$, forming $H_{6 e_1}$ instead of $H_{6 e_2}$.
\end{itemize}

The updated node-hyperedge relationships after applying the operator are shown in the right of Fig.~\ref{initialization-operator}. The newly generated hypergraph, along with the original hypergraph, collectively forms the initial population. This ensures structural diversity for subsequent optimization processes.

\subsection{Crossover Operator}

\begin{figure}[h]
  \begin{center}
  \includegraphics[width=3in]{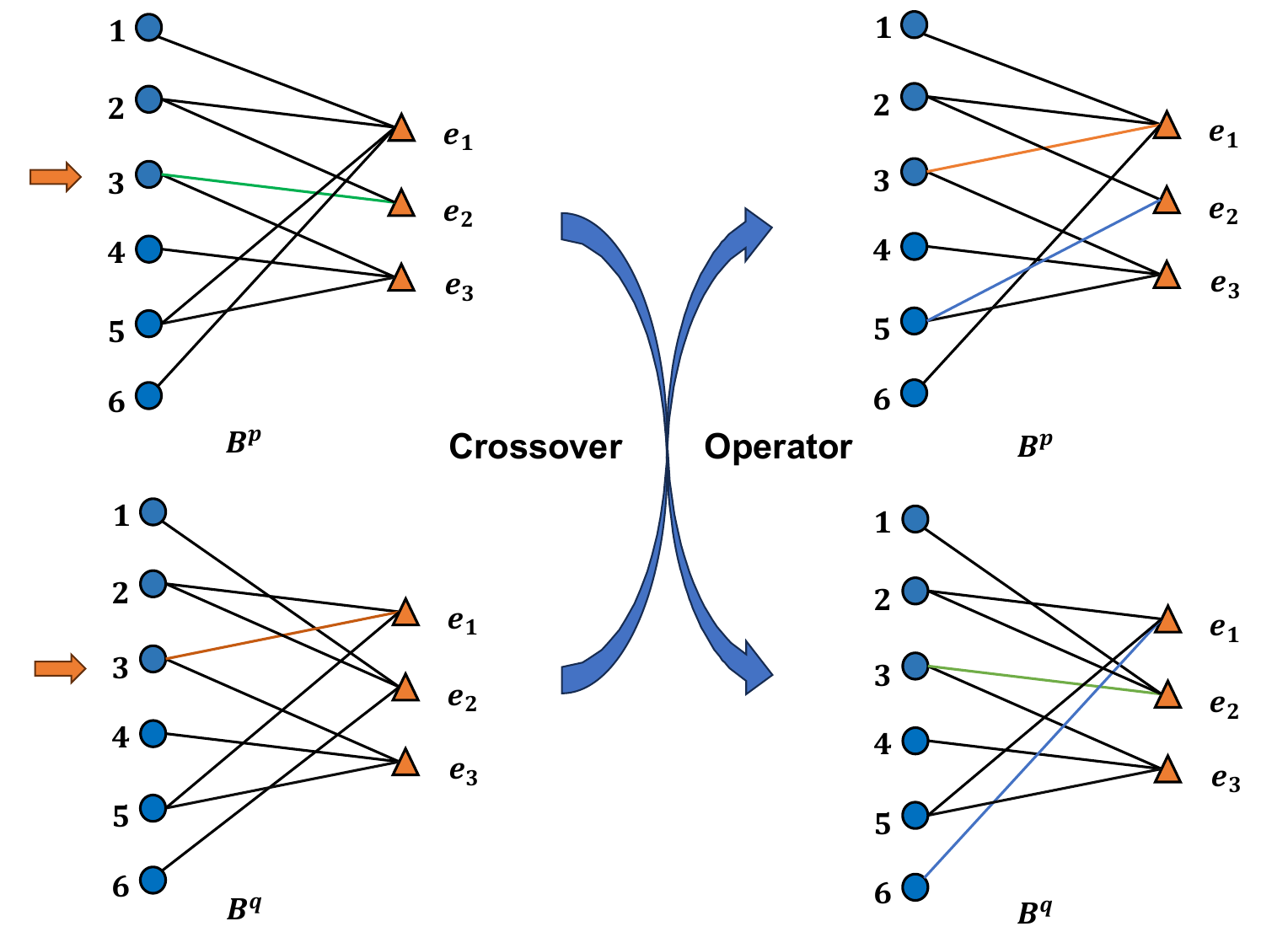}\\
  \caption{The illustration of the crossover operator.}\label{Crossover-operator}
  \end{center}
\end{figure}

The uniform crossover operator is designed based on a bipartite graph representation, utilizing the structural properties of hypergraphs to exchange node-hyperedge relationships between two bipartite graphs $B^p$ and $B^q$.  
Fig.~\ref{Crossover-operator} takes an example to illustrate how the crossover operator works. In this example, a hypergraph, represented as a bipartite graph, consists of six nodes and three hyperedges. The node-hyperedge relationships of $B^p$ and $B^q$ are shown in the top-left and bottom-left of Fig.~\ref{Crossover-operator}, respectively. Supposing that node $3$ is selected as a crossover point, we identify the distinct node-hyperedge relationships associated with node $3$ in $B^p$ and $B^q$. This specifically forms $H_{3 e_2}$ and $H_{2 e_1}$, subsequently performing the swap operation as follows:  

\begin{itemize}  
\item Step 1: For $B^p$, we first locate node $5$, which is connected to $e_1$ but not to $e_2$. Then, the swap operator $\mathcal{S}$ is then applied to modify the node-hyperedge relationships:  

   \begin{equation}  
   \mathcal{S}({H_{3 e_2}}, {H_{5 e_1}}) = ({H_{3 e_1}}, {H_{5 e_2}})  
   \end{equation}  

Following this operation, the node-hyperedge relationship $H_{3 e_1}$ is successfully transferred to $B^p$.

\item Step 2: For $B^q$, we first locate node $6$, which is connected to $e_2$ but not to $e_1$. Afterwards, the swap operator $\mathcal{S}$ is then applied to modify the node-hyperedge relationships: 

    \begin{equation}
   \mathcal{S}({H_{3 e_1}}, {H_{6 e_2}}) = ({H_{2 e_2}}, {H_{6 e_1}})
    \end{equation}

After this step, the node-hyperedge relationship $H_{3 e_2}$ is updated to $B^q$.  

\end{itemize}

The updated node-hyperedge relationships after applying the crossover operator are shown in the right of Fig.~\ref{Crossover-operator}. This process preserves both the node hyperdegree and hyperedge cardinality while facilitating structural diversity in the population.

\subsection{Mutation Operator}

\begin{figure}[h]
  \begin{center}
  \includegraphics[width=5in]{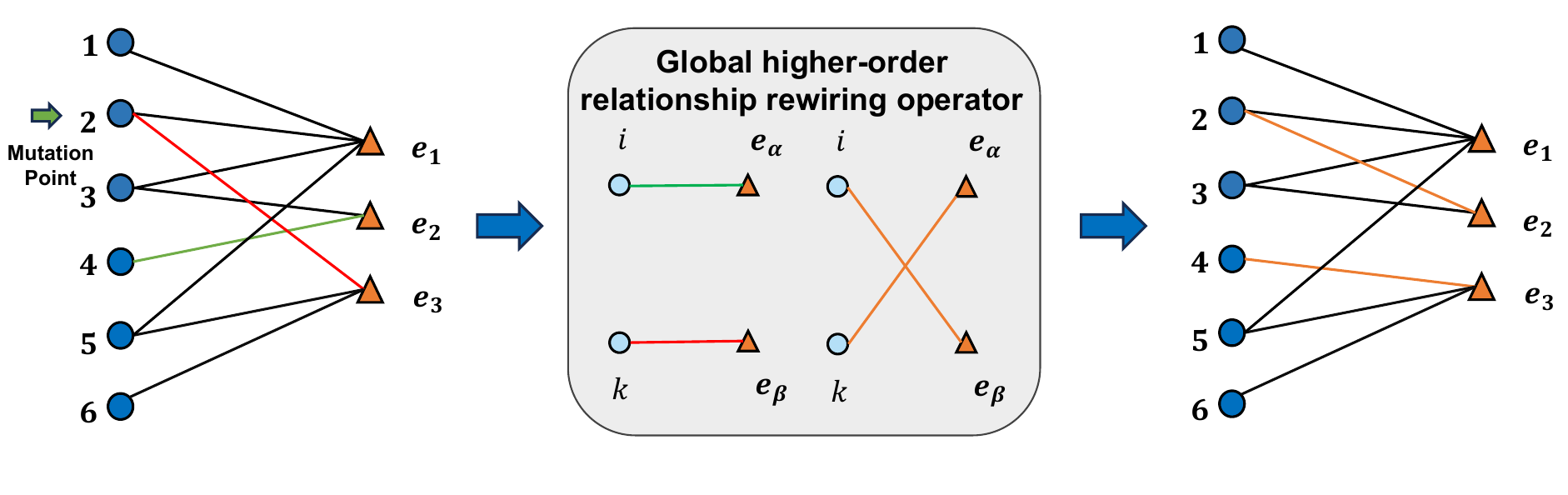}\\
  \caption{The illustration of the mutation operator.}\label{Mutation-operator}
  \end{center}
\end{figure}

The multi-point mutation operator is designed to enhance robustness by prioritizing the mutation of key nodes. 
Fig.~\ref{Mutation-operator} illustrates an example of how a parent individual (i.e, hypergraph) is mutated to a new offspring. In this example, the parent individual on the left consists of six nodes and three hyperedges. Supposing that node $2$ is identified as a mutation point based on its hyperdegree ranking, the mutation operator then modifies the node-hyperedge relationships as follows:

\begin{itemize}  
\item Step 1: A node-hyperedge relationship $H_{2 e_3}$ associated with node $2$ is randomly selected.  
\item Step 2: Another node-hyperedge relationship $H_{4 e_2}$ is randomly chosen, ensuring that $2 \notin e_{2}$ and $4 \notin e_{3}$.
\item Step 3: The global higher-order relationship rewiring operator $\mathcal{G}$ is then applied to swap the selected relationships, leading to the following update:  

   \begin{equation}  
   \mathcal{G}({H_{2 e_{3}}}, {H_{4 e_{2}}}) = ({H_{2 e_{2}}}, {H_{4 e_{3}}}).  
   \end{equation}  

\end{itemize}  

Following this operation, the updated node-hyperedge relationships are depicted in the right of Fig.~\ref{Mutation-operator}.

\section{Experiment Design}

\subsection{Hypergraph Configurations}

In the experiments on synthetic hypergraphs, we selected two types of hypergraphs: ER hypergraphs and SF hypergraphs. All the experiments were conducted on their largest connected component.

\subsubsection{ER Hypergraphs}

The Erdős-Rényi (ER) hypergraph model extends the classical Erdős-Rényi random graph model to higher-order networks. In this study, we employ an ER hypergraph generation method based on bipartite graphs, as implemented in the HyperNetX package and originally proposed by Aksoy et al.. This method constructs a hypergraph by first defining a bipartite network consisting of two disjoint sets: a set of $N$ nodes and a set of $M$ hyperedges. Each node $i$ is stochastically assigned to a hyperedge $e_\gamma$ with a probability $p$, thus determining the incidence matrix $H$. This model assumes that each node-hyperedge relationship $H_{ie_\gamma}$ is connected independently with probability $p$, leading to an expected number of node-hyperedge relationship given by

\begin{equation} {\cal E}\left[ {\left| E \right|} \right] = p \cdot N \cdot M, \end{equation}
where ${\cal E}\left[ {\left| E \right|} \right]$ denotes the expect number of node-hyperedge connections in the generated hypergraph. The core of the algorithm involves iterating over each node and determining its connections to hyperedges through a stochastic sampling process. Instead of evaluating each possible connection independently, which would be computationally expensive for large networks, the algorithm employs an optimized sampling method based on logarithmic transformation. Specifically, for a given node $i$, a random number $r \in (0,1)$ is drawn, and the next hyperedge index $e_\gamma^t$ to which $i$ is connected is determined as

\begin{equation} e_\gamma ^t = e_\gamma ^{t - 1} + \left\lfloor {\frac{{\log r}}{{\log (1 - p)}}} \right\rfloor, \end{equation}
where $e^0_\gamma =0$. This allows the model to efficiently sample incident pairs without redundant iteration over all possible connections. If the sampled hyperedge index 
$e^t_\gamma$ remains within the valid range, the node-hyperedge connection $H_{ie_\gamma^t}$ is recorded. Once all node-hyperedge pairs are determined, the resulting bipartite graphs can be converted into hypergraphs. Note that the average hyperdegree is $\left\langle k \right\rangle = p \cdot M$.

\subsubsection{SF Hypergraphs}
The Scale-Free (SF) hypergraph model generalizes the classical Scale-Free (SF) graph model to higher-order networks, where both the hyperdegree of nodes and the cardinality of hyperedges follow a power-law distribution.  In this work, we implement a scale-free hypergraph generation method based on the Chung-Lu (CL) model, which constructs a hypergraph by assigning connections probabilistically based on a predefined node degree distribution.

The generation process begins with two input parameters. 
The first is the hyperedge cardinality sequence $\{m_\gamma\}$, where each hyperedge $e_\gamma$ has a specified cardinality $m_\gamma$. The second is the node degree sequence $\{k_i\}$, which determines the number of hyperedges each node is associated with. Both sequences follow a power-law distribution
\begin{equation} 
\begin{array}{l}
P({k_i})\sim k_i^{ - \lambda },\\
P({m_\gamma })\sim m_\gamma ^{ - \lambda },
\end{array}
\end{equation} 
where $\lambda$ is the exponent governing the distribution, ensuring that a few nodes (hyperedges) have significantly higher hyperdegrees than the majority.  

To construct the hypergraph, we first compute the probability of each node being selected for inclusion in a hyperedge. This probability is determined by normalizing the node degree sequence, yielding  

\begin{equation} 
p_i = \frac{k_i}{\sum_{i} k_i},  
\end{equation} 
where $p_i$ represents the probability of selecting node $i$. 

Using this probability distribution, we create a cumulative probability range list $R = \{R_0, R_1, ..., R_N\}$, where  

\begin{equation} 
R_i = \sum_{j=0}^{i-1} p_j, \quad R_0 = 0, \quad R_N = 1.
\end{equation} 

This cumulative distribution enables efficient node selection. Specifically, for each hyperedge $e_\gamma$ with predefined size $m_\gamma$, we randomly generate a number $r \in (0,1)$ and determine the corresponding node $i$ by locating the smallest index $j$ such that $R_j \leq r < R_{j+1}$. If the selected node has already been assigned to the hyperedge, the process repeats until the hyperedge reaches the required size. This procedure is repeated for all the hyperedges until the entire hypergraph structure is generated.

The developed hypergraph maintains a power-law hyperdegree distribution while ensuring that hyperedges are formed according to the predefined size sequence. The Chung-Lu-based approach enables scalable and computationally efficient generation of large-scale scale-free hypergraphs. This makes it suitable for modeling real-world systems where heterogeneous connectivity patterns are crucial, such as biological networks, social systems, and communication infrastructures. Note that the Chung-Lu-based method requires the total sums of the sequences $\{m_\gamma\}$ and $\{k_i\}$ to be approximately equal. To achieve this balance, we employ an iterative method that dynamically adjusts the upper bound $m_{\max}$ of the power-law distribution.

\subsection{Parameter Settings}

The Meta‑HAROF framework was implemented using the DEAP package in Python. During the evolutionary process, we adopted an elitism strategy, where the top-performing individual from the previous generation is directly copied to the current generation to preserve high-quality solutions. The detailed parameter settings of the Meta‑HAROF framework are summarized in Table \ref{TB1}.  

To comprehensively evaluate the framework’s performance, experiments were conducted on 16 hypergraphs with diverse structural characteristics. The hypergraphs were categorized into two types: ER and SF, each constructed with different configurations. Specifically, we considered two network scales ($N = 1,000$ and $N = 2,000$), two node-to-hyperedge ratios ($2:1$ and $1:1$), and distinct structural parameters for each type. For the ER hypergraphs, the average hyperdegree $\left\langle k \right\rangle$ was set to $2$ and $3$, following the naming convention ER-N-M-$\left\langle k \right\rangle$, where $N$ is the number of nodes, $M$ is the number of hyperedges, and $\left\langle k \right\rangle$ denotes the average hyperdegree. For the SF hypergraphs, the power-law exponent $\lambda$ was set to $1.5$ and $2$, adhering to the naming convention SF-N-M-$\lambda$, where $N$ represents the number of nodes, $M$ the number of hyperedges, and $\lambda$ the power-law exponent. These configurations ensure a diverse and representative set of hypergraphs, allowing a thorough assessment of the effectiveness of the Meta‑HAROF framework in different network structures.

All the experiments were conducted on a server equipped with a 4th generation Intel Xeon Scalable processor (3.10 GHz) and 256 GB of RAM, ensuring sufficient computational resources for evaluating the framework under different hypergraph configurations. 
\begin{table}
\centering
\caption{The parameter settings of G-CIIM.}
\begin{tabular}{cc}
\hline
Parameter&  Value\\
\hline
The size of population& 512\\
Crossover probability & 0.9\\
The proportion of crossover nodes & 0.1\\
Mutation probability & 0.1\\
The proportion of mutated nodes & 0.005\\
The number of generations& 800\\
Tournament size& 5\\
\hline
\end{tabular}
\label{TB1}
\end{table}

\subsection{Network Attacks}
In the experiments, we have considered four types of network attacks, introduced as follows:

\textbf{Malicious attacks:} A malicious attack is a network disruption strategy that prioritizes the removal of nodes based on structural importance, typically determined by centrality measures such as degree, betweenness centrality, or eigenvector centrality. By preferentially eliminating the most critical nodes, this attack rapidly compromises network connectivity. In this study, we adopt hyperdegree as the centrality measure and preferentially remove nodes with higher hyperdegree.

\textbf{Random attacks:} A random attack is an untargeted disruption strategy where nodes are removed randomly, modeling failures caused by natural factors such as aging, weather conditions, or other stochastic events.

\textbf{Accumulated attacks:} An accumulated attack is an adaptive disruption strategy where node centrality metrics are recalculated after each removal, ensuring that the most central node is always targeted. This attack approach simulates an intelligent adversary that continuously identifies and eliminates key nodes, accelerating network disintegration. In this study, we use hyperdegree as the centrality measure.

\textbf{Hyperedge attacks:} A hyperedge attack is a unique hypergraph disruption strategy, targeting hyperedges as fundamental relational units connecting multiple nodes. This attack approach seeks to compromise the structural integrity of the hypergraph by identifying and removing the most critical hyperedges. In this study, we prioritize the removal of hyperedges with the largest cardinality.

\section{Additional Experiment Results}

\subsection{Node Attacks}

\begin{table}[h]
    \centering
    \sisetup{table-number-alignment = center, round-mode = places, round-precision = 3}
    \caption{The average robustness comparison of original and optimized hypergraphs under malicious attack. The average robustness (\textbf{Ave. $R_{N_{C}}$}), standard deviation (\textbf{Std. $R_{N_{C}}$}) and Significance test (\textbf{Sig.}) are reported for both original and optimized hypergraphs. In the significance test, "$-$" denotes that the robustness of the optimized hypergraph is significantly lower than that of the original hypergraphs, a "$+$" indicates superior robustness of the optimized hypergraph, while "$=$" signifies no statistically significant difference between the two.}
    \label{tab:robustness_comparison}
    \begin{tabular}{l S[table-format=1.3] S[table-format=1.3] S[table-format=1.3] S[table-format=1.3] l}
        \toprule
        \textbf{Hypergraph Name} & \textbf{Ave. $R^{Orig}_{N_{T}}$} & \textbf{Std. $R^{Orig}_{N_{T}}$} & \textbf{Ave. $R^{Opt}_{N_{T}}$} & \textbf{Std. $R^{Opt}_{N_{T}}$} 
 & \textbf{Sig.} \\
        \midrule
        ER-1000-1000-2  & 0.182 & 0.012 & \textbf{0.308} & 0.017 & $+$\\
        ER-1000-1000-3  & 0.329 & 0.008 & \textbf{0.423} & 0.018 & $+$\\
        ER-1000-500-2   & 0.285 & 0.011 & \textbf{0.376} & 0.009 & $+$\\
        ER-1000-500-3   & 0.400 & 0.005 & \textbf{0.470} & 0.004 & $+$\\
        ER-2000-2000-2  & 0.181 & 0.009 & \textbf{0.276} & 0.012 & $+$\\
        ER-2000-2000-3  & 0.329 & 0.004 & \textbf{0.397} & 0.008 & $+$\\
        ER-2000-1000-2  & 0.280 & 0.007 & \textbf{0.372} & 0.005 & $+$\\
        ER-2000-1000-3  & 0.397 & 0.003 & \textbf{0.463} & 0.004 & $+$\\
        SF-1000-1000--1.5  & 0.336 & 0.005 & \textbf{0.423} & 0.029 & $+$\\
        SF-1000-1000--2    & 0.255 & 0.007 & \textbf{0.397} & 0.019 & $+$\\
        SF-1000-500--1.5   & 0.329 & 0.009 & \textbf{0.412} & 0.028 & $+$\\
        SF-1000-500--2     & 0.279 & 0.007 & \textbf{0.373} & 0.004 & $+$\\
        SF-2000-2000--1.5  & 0.331 & 0.008 & \textbf{0.397} & 0.017 & $+$\\
        SF-2000-2000--2    & 0.245 & 0.007 & \textbf{0.376} & 0.005 & $+$\\
        SF-2000-1000--1.5  & 0.329 & 0.004 & \textbf{0.407} & 0.019 & $+$\\
        SF-2000-1000--2    & 0.274 & 0.006 & \textbf{0.369} & 0.004 & $+$\\
        \bottomrule
    \end{tabular}
\end{table}

Table \ref{tab:robustness_comparison} presents a detailed comparison of the robustness metric $R_{N_M}$ between the original and optimized hypergraphs under malicious attacks. These results demonstrate a consistent improvement, with $R_{N_M}$ increasing by 16.6\% to 68.7\% after optimization. This confirms the effectiveness of our approach in enhancing hypergraph robustness.  

\begin{figure}[h]
  \begin{center}
  \includegraphics[width=5in]{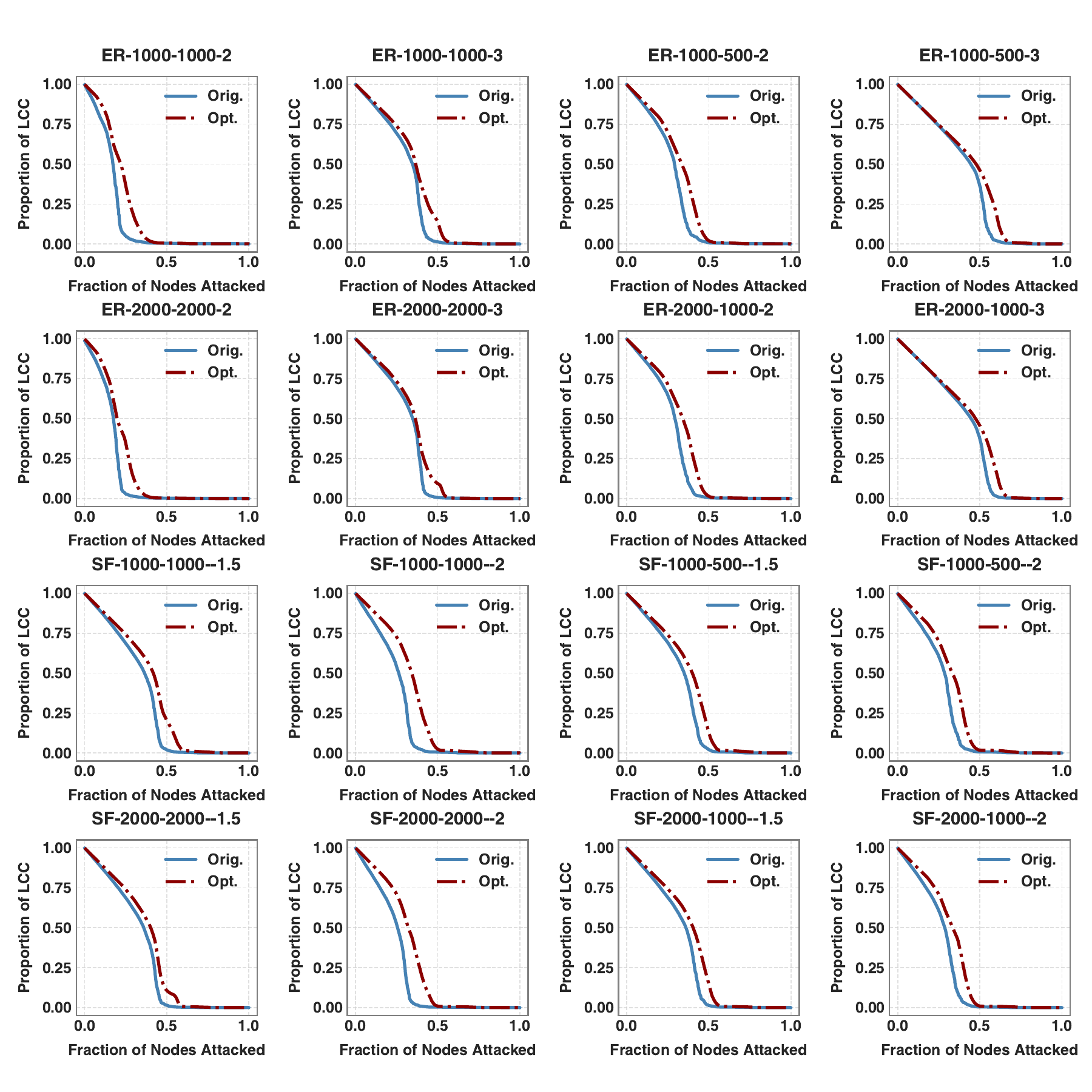}\\
  \caption{\textbf{Percolation curves of synthetic hypergraphs under cumulative attacks.} Each subplot illustrates the evolution of the LCC as the fraction of nodes attacked, under specific structural parameter settings, for both the original and optimized hypergraphs.}\label{synthetic networks Cumulative}
  \end{center}
\end{figure}

To further evaluate the robustness of the optimized hypergraphs under different attack strategies, we extend our analysis to cumulative and random attacks. The percolation curves and robustness metric $R_{N_C}$ under the cumulative attacks are presented in Fig. \ref{synthetic networks Cumulative} and Table \ref{tab:robustness_comparison_cumulative}, respectively. As shown in Fig. \ref{synthetic networks Cumulative}, the optimized hypergraphs exhibit significantly larger percolation curve areas compared to the original hypergraphs. Furthermore, Table \ref{tab:robustness_comparison_cumulative} indicates that the robustness metric $R^{Opt}_{N_{C}}$ of the optimized hypergraphs improves by 8\% to 35.3\% relative to the original hypergraphs. These results demonstrate that hypergraphs optimized against malicious attacks also exhibit enhanced robustness under the cumulative attacks.

\begin{table}[h]
    \centering
    \sisetup{table-number-alignment = center, round-mode = places, round-precision = 3}
    \caption{The average robustness comparison of original and optimized hypergraphs under cumulative attack. The average robustness (\textbf{Ave. $R_{N_{C}}$}), standard deviation (\textbf{Std. $R_{N_{C}}$}) and Significance test (\textbf{Sig.}) are reported for both original and optimized hypergraphs. In the significance test, "$-$" denotes that the robustness of the optimized hypergraph is significantly lower than that of the original hypergraphs, a "$+$" indicates superior robustness of the optimized hypergraph, while "$=$" signifies no statistically significant difference between the two.}
    \label{tab:robustness_comparison_cumulative}
    \begin{tabular}{l S[table-format=1.3] S[table-format=1.3] S[table-format=1.3] S[table-format=1.3] l}
        \toprule
        \textbf{Hypergraph Name} & \textbf{Ave. $R^{Orig}_{N_{C}}$} & \textbf{Std. $R^{Orig}_{N_{C}}$} & \textbf{Ave. $R^{Opt}_{N_{C}}$} & \textbf{Std. $R^{Opt}_{N_{C}}$} & \textbf{Sig.}\\
        \midrule
        ER-1000-1000-2  & 0.164 & 0.010 & \textbf{0.221} & 0.011 & $+$\\
        ER-1000-1000-3  & 0.299 & 0.008 & \textbf{0.339} & 0.012 & $+$\\
        ER-1000-500-2   & 0.270 & 0.012 & \textbf{0.319} & 0.007 & $+$\\
        ER-1000-500-3   & 0.383 & 0.004 & \textbf{0.417} & 0.003 & $+$\\
        ER-2000-2000-2  & 0.161 & 0.007 & \textbf{0.209} & 0.009 & $+$\\
        ER-2000-2000-3  & 0.298 & 0.004 & \textbf{0.328} & 0.006 & $+$\\
        ER-2000-1000-2  & 0.267 & 0.007 & \textbf{0.313} & 0.005 & $+$\\
        ER-2000-1000-3  & 0.383 & 0.004 & \textbf{0.414} & 0.004 & $+$\\
        SF-1000-1000--1.5  & 0.319 & 0.005 & \textbf{0.371} & 0.017 & $+$\\
        SF-1000-1000--2    & 0.235 & 0.006 & \textbf{0.318} & 0.012 & $+$\\
        SF-1000-500--1.5   & 0.311 & 0.010 & \textbf{0.358} & 0.016 & $+$\\
        SF-1000-500--2     & 0.259 & 0.007 & \textbf{0.314} & 0.006 & $+$\\
        SF-2000-2000--1.5  & 0.316 & 0.007 & \textbf{0.355} & 0.011 & $+$\\
        SF-2000-2000--2    & 0.227 & 0.006 & \textbf{0.305} & 0.005 & $+$\\
        SF-2000-1000--1.5  & 0.313 & 0.005 & \textbf{0.359} & 0.010 & $+$\\
        SF-2000-1000--2    & 0.257 & 0.004 & \textbf{0.310} & 0.004 & $+$\\
        \bottomrule
    \end{tabular}
\end{table}

\begin{figure}[h]
  \begin{center}
  \includegraphics[width=5in]{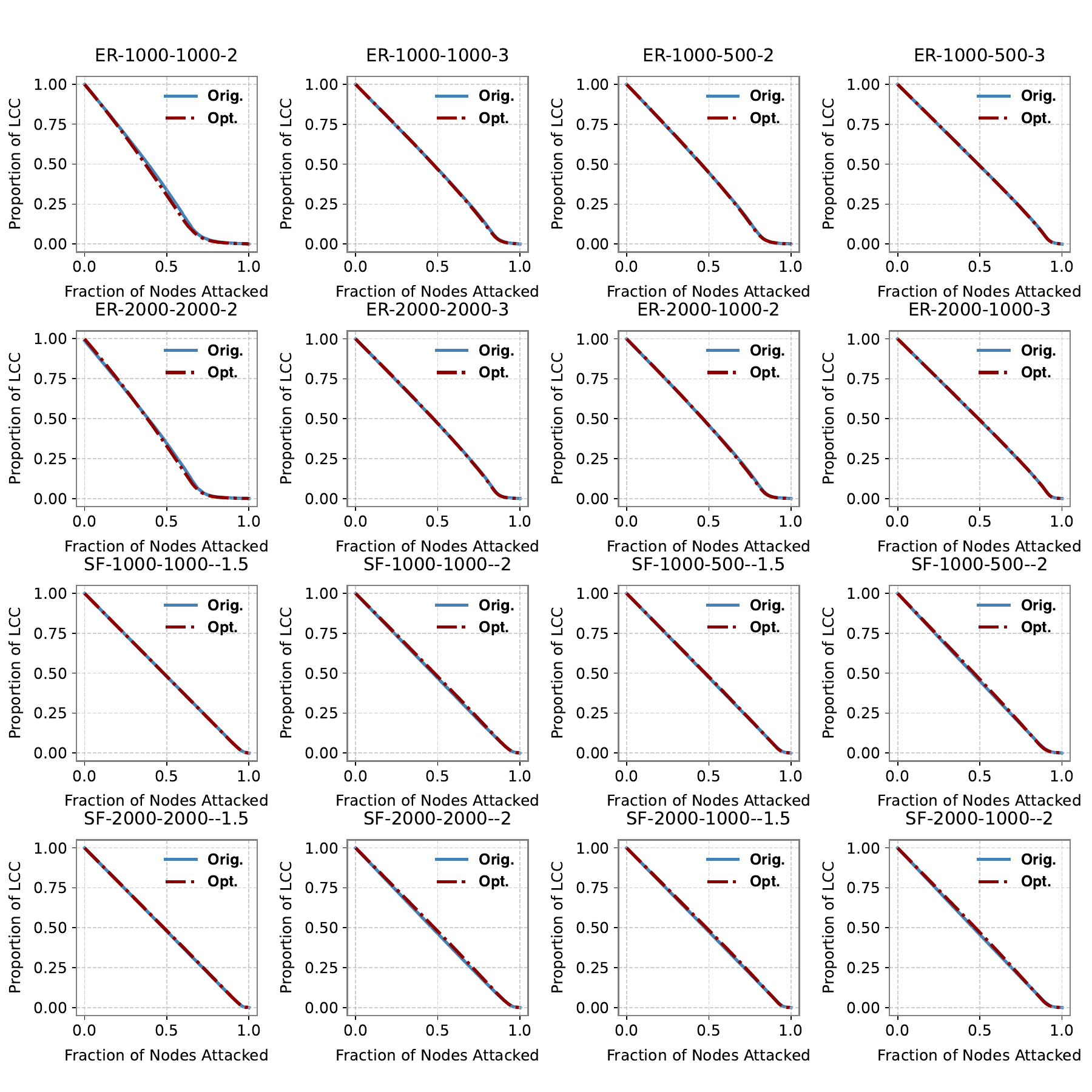}\\
  \caption{\textbf{Percolation curves of synthetic hypergraphs under random attacks.} Each subplot illustrates the evolution of the LCC as the fraction of nodes attacked, under specific structural parameter settings, for both the original and optimized hypergraphs.}\label{synthetic networks random}
  \end{center}
\end{figure}

The simulation results of the original and optimized hypergraphs under the random attacks are presented in Fig. \ref{synthetic networks random} and Table \ref{tab:robustness_comparison_random}. As shown in the experiments, the robustness metric $R_{N_R}$ exhibits no significant change between the optimized and original hypergraphs. This indicates that the Meta‑HAROF framework enhances robustness against malicious attacks without compromising the ability against random attacks.

\begin{table}[h]
    \centering
    \sisetup{table-number-alignment = center, round-mode = places, round-precision = 3}
    \caption{The average robustness comparison of original and optimized hypergraphs under random attack. The average robustness (\textbf{Ave. $R_{N_{C}}$}), standard deviation (\textbf{Std. $R_{N_{C}}$}) and Significance test (\textbf{Sig.}) are reported for both original and optimized hypergraphs. In the significance test, "$-$" denotes that the robustness of the optimized hypergraph is significantly lower than that of the original hypergraphs, a "$+$" indicates superior robustness of the optimized hypergraph, while "$=$" signifies no statistically significant difference between the two.}
    \label{tab:robustness_comparison_random}
    \begin{tabular}{l S[table-format=1.3] S[table-format=1.3] S[table-format=1.3] S[table-format=1.3] l}
        \toprule
        \textbf{Hypergraph Name} & \textbf{Ave. $R^{Orig}_{N_{C}}$} & \textbf{Std. $R^{Orig}_{N_{C}}$} & \textbf{Ave. $R^{Opt}_{N_{C}}$} & \textbf{Std. $R^{Opt}_{N_{C}}$} & \textbf{Sig.}\\
        \midrule
        ER-1000-1000-2  & \textbf{0.392} & 0.006 & 0.380 & 0.007 & $-$\\
        ER-1000-1000-3  & 0.466 & 0.004 & 0.465 & 0.003 & $=$\\
        ER-1000-500-2   & 0.460 & 0.004 & 0.460 & 0.005 & $=$\\
        ER-1000-500-3   & 0.489 & 0.001 & 0.489 & 0.001 & $=$\\
        ER-2000-2000-2  & 0.390 & 0.007 & 0.387 & 0.006 & $=$\\
        ER-2000-2000-3  & 0.467 & 0.002 & 0.467 & 0.002 & $=$\\
        ER-2000-1000-2  & 0.461 & 0.001 & 0.460 & 0.003 & $=$\\
        ER-2000-1000-3  & 0.489 & 0.001 & \textbf{0.490} & 0.001 & $+$\\
        SF-1000-1000--1.5  & 0.488 & 0.001 & \textbf{0.489} & 0.004 & $+$\\
        SF-1000-1000--2    & 0.478 & 0.002 & \textbf{0.487} & 0.002 & $+$\\
        SF-1000-500--1.5   & 0.483 & 0.002 & \textbf{0.485} & 0.003 & $+$\\
        SF-1000-500--2     & 0.470 & 0.003 & \textbf{0.476} & 0.003 & $+$\\
        SF-2000-2000--1.5  & 0.488 & 0.001 & \textbf{0.491} & 0.002 & $+$\\
        SF-2000-2000--2    & 0.477 & 0.001 & \textbf{0.487} & 0.001 & $+$\\
        SF-2000-1000--1.5  & 0.483 & 0.001 & \textbf{0.488} & 0.001 & $+$\\
        SF-2000-1000--2    & 0.470 & 0.002 & \textbf{0.477} & 0.002 & $+$\\
        \bottomrule
    \end{tabular}
\end{table}

\subsection{Hyperedge Attack}

\begin{figure}[h]
  \begin{center}
  \includegraphics[width=5in]{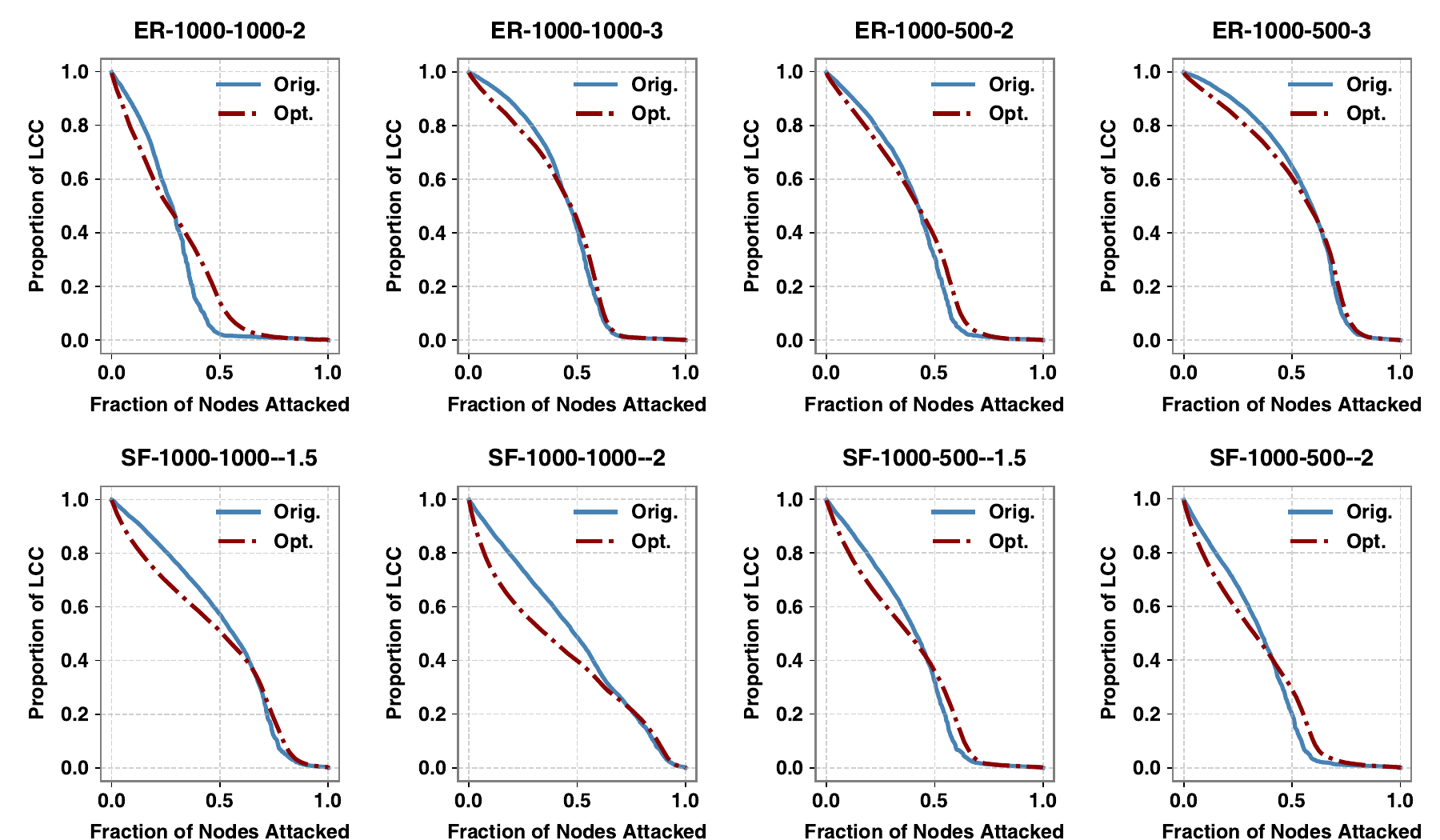}\\
  \caption{Percolation curves of synthetic hypergraphs under hyperedges attack. Each subplot depicts how the fraction of nodes in the LCC evolves as the fraction of hyperedges attacked for a specific structure parameter setting in the original and optimized hypergraphs. }\label{synthetic networks hyperedge}
  \end{center}
\end{figure}

To validate the generalizability of the Meta‑HAROF framework, we optimized hypergraphs with varying structural parameters to enhance robustness against the hyperedge attacks. The experimental results are presented in Fig. \ref{synthetic networks hyperedge} and Table \ref{tab:robustness_comparison_hyperedge}. As shown in Fig. \ref{synthetic networks hyperedge}, the percolation curve area of the optimized hypergraphs is significantly larger than that of the original hypergraphs. Moreover, Table \ref{tab:robustness_comparison_hyperedge} indicates that the robustness metric $R^{Opt}_{M_{H}}$ improves by 18.2\% to 45.6\% compared to the original hypergraphs. These findings demonstrate that the Meta‑HAROF framework effectively enhances robustness against both the node and hyperedge attacks, confirming its generalizability.

\begin{table}[h]
    \centering
    \sisetup{table-number-alignment = center, round-mode = places, round-precision = 3}
    \caption{The average robustness comparison of original and optimized hypergraphs under hyperedge attack. The average robustness (\textbf{Ave. $R_{M_{H}}$}), standard deviation (\textbf{Std. $R_{M_{H}}$}) and Significance test (\textbf{Sig.}) are reported for both original and optimized hypergraphs. In the significance test, "$-$" denotes that the robustness of the optimized hypergraph is significantly lower than that of the original hypergraphs, a "$+$" indicates superior robustness of the optimized hypergraph, while "$=$" signifies no statistically significant difference between the two.}
    \label{tab:robustness_comparison_hyperedge}
    \begin{tabular}{l S[table-format=1.3] S[table-format=1.3] S[table-format=1.3] S[table-format=1.3] l}
        \toprule
        \textbf{Hypergraph Name} & \textbf{Ave. $R^{Orig}_{M_{H}}$} & \textbf{Std. $R^{Orig}_{M_{H}}$} & \textbf{Ave. $R^{Opt}_{M_{H}}$} & \textbf{Std. $R^{Opt}_{M_{H}}$} & \textbf{Sig.}\\
        \midrule
        ER-1000-1000-2  & 0.284 & 0.012 & \textbf{0.413} & 0.012 & $+$\\
        ER-1000-1000-3  & 0.441 & 0.014 & \textbf{0.548} & 0.013 & $+$\\
        ER-1000-500-2   & 0.392 & 0.015 & \textbf{0.489} & 0.014 & $+$\\
        ER-1000-500-3   & 0.539 & 0.009 & \textbf{0.637} & 0.009 & $+$\\
        SF-1000-1000--1.5  & 0.504 & 0.006 & \textbf{0.644} & 0.006 & $+$\\
        SF-1000-1000--2    & 0.476 & 0.007 & \textbf{0.596} & 0.006 & $+$\\
        SF-1000-500--1.5   & 0.380 & 0.014 & \textbf{0.481} & 0.026 & $+$\\
        SF-1000-500--2     & 0.338 & 0.012 & \textbf{0.425} & 0.018 & $+$\\
        \bottomrule
    \end{tabular}
\end{table}

\section{Robust Hypergraph Generation Method}

To validate the empirical observation that nodes with higher average hyperdegrees tend to connect to hyperedges with lower cardinalities in optimized hypergraphs, we propose a robust hypergraph generation method based on power law bias and weighted sampling without replacement. This method generates hypergraphs, where both the node hyperdegree and hyperedge cardinality follow a distribution. Note that it is typically not limited to a specific distribution.   

Similar to the SF hypergraph generation process, the method begins with two input parameters: the hyperedge cardinality sequence $\{m_\gamma\}$ and the node hyperdegree sequence $\{k_i\}$. Initially, a global allocation table, named $remaining$, is established, where each node $i$ is assigned an initial quota corresponding to its target hyperdegree, i.e., $remaining[i]=k_i$. The hyperedges are then sorted in ascending order of cardinality to prioritize the allocation of smaller hyperedges first. This sorting ensures that hyperedges with smaller cardinality are prioritized in node allocation. For each hyperedge of cardinality $k$, a node selection process is conducted using a weighted sampling method without replacement. The sampling weight of each candidate node $i$ is determined by its original hyperdegree $k_i$ raised to a power-law bias factor $\beta_\gamma$:

\begin{equation}
w_i = k_i^{\beta_\gamma},
\end{equation}
where $\beta_\gamma$ is dynamically adjusted based on the current hyperedge cardinality:

\begin{equation}
\beta_\gamma = \frac{\max(\{m_\gamma\}) }{m_\gamma} \cdot rb,
\end{equation}
where $rb$ is a tunable robustness parameter that controls the degree of preferential attachment, and $\frac{\max(\{m_\gamma\}) }{m_\gamma}$ ensures that hyperedges with smaller cardinality correspond to larger values of $\beta_\gamma $. For $rb > 1$, high-hyperdegree nodes preferentially connect to hyperedges of smaller cardinality, whereas for $rb < -1$, low-hyperdegree nodes exhibit a similar preference for connecting to hyperedges with smaller cardinality. The sampling procedure is performed iteratively: a candidate node is selected with probability proportional to its weight; once chosen, its remaining quota in $remaining$ is decremented to enforce a hard constraint on node participation. As the process continues, each hyperedge is sequentially populated until all hyperedges receive their designated nodes while respecting the hyperdegree constraints. The generation method of RB hypergraphs is detailed in Algorithm~\ref{alg:hyperRB}.

\begin{algorithm}[htb]
\caption{Robust Hypergraph Generation Method}
\label{alg:hyperRB}
\begin{algorithmic}[1]
\Require $hpe\_size$ (list of hyperedge cardinalities), $hnd\_deg$ (list of node hyperdegrees), $rb$ (robustness scaling factor)
\Ensure $edge\_aff\_dict$ (mapping of hyperedges to selected nodes)

\State Initialize remaining quota for each node: $remaining[i] \gets hnd\_deg[i], \forall i$
\State Sort hyperedges in ascending order of cardinality
\State $max\_cardinality \gets \max(hpe\_size)$

\For{each hyperedge $(edge\_index, size)$ in sorted hyperedges}
    \State Compute degree bias: $\beta \gets \frac{max\_cardinality}{size} \cdot rb$
    \State Initialize empty set $selected\_nodes$
    
    \For{$t = 1$ to $size$}
        \State Compute weighted probability for each candidate node:
        \[
        w_i = (hnd\_deg[i])^\beta, \quad \forall i \in remaining \text{ with } remaining[i] > 0
        \]
        \State Normalize weights and sample a node $i$ using weighted probability
        \State Add $i$ to $selected\_nodes$ and decrement $remaining[i]$
    \EndFor

    \If{$|selected\_nodes| < size$}
        \State \textbf{break}
    \EndIf

    \State Assign $selected\_nodes$ to hyperedge $edge\_index$
\EndFor
\State \Return $edge\_aff\_dict$
\end{algorithmic}
\end{algorithm}


